\documentclass[a4paper,11pt]{article}
\pdfoutput=1 

\usepackage{jheppub} 
\usepackage{amsmath, amsfonts}

\usepackage[T1]{fontenc} 
\usepackage{xcolor}
\usepackage{subfigure}
\usepackage{stmaryrd}
\usepackage{enumerate}
\usepackage{tikz}
\usetikzlibrary{calc,intersections,positioning,arrows}

\newcommand{\beq}{\begin{equation}}
\newcommand{\eeq}{\end{equation}}
\newcommand{\bea}{\begin{eqnarray}}
\newcommand{\eea}{\end{eqnarray}}
\newcommand{\al}{\alpha}

\newcommand{\der}{\partial}

\newcommand{\N}{{\cal{N}}}

\newcommand{\textoverline}[1]{$\overline{\mbox{#1}}$}
\newcommand{\MKK}{M_{\mathrm{KK}}}

\usepackage{color}

\usepackage[normalem]{ulem}  

\newcommand{\non}{\nonumber\\}

\newcommand{\be}{\begin{equation}}
\newcommand{\ee}{\end{equation}}

\title{\boldmath Holographic quarkyonic matter}

\author[a]{Nicolas Kovensky}
\author[a]{and Andreas Schmitt}

\affiliation[a]{Mathematical Sciences and STAG Research Centre, University of Southampton, Highfield Campus, Southampton
SO17 1BJ, United Kingdom.}

\emailAdd{n.kovensky@soton.ac.uk}
\emailAdd{a.schmitt@soton.ac.uk}

\abstract{We point out a new configuration in the Witten-Sakai-Sugimoto model, allowing baryons in the pointlike approximation to coexist with fundamental quarks. The resulting phase is a holographic realization of quarkyonic matter, which is predicted to occur in QCD at a large number of colors, and possibly plays a role in real-world QCD as well. We find that holographic quarkyonic matter is chirally symmetric and that,  for large baryon chemical potentials, it is energetically preferred over pure nuclear matter and over pure quark matter. The zero-temperature transition from nuclear matter to the quarkyonic phase is of first order in the chiral limit and for a realistic pion mass. 
For pion masses far beyond the physical point we observe a quark-hadron continuity due to the presence of quarkyonic matter.}

\begin{document} 
\maketitle
\flushbottom

\section{Introduction}
\label{sec:intro}

The phase structure of Quantum Chromodynamics (QCD) at large, but not asymptotically large, baryon densities is poorly understood. Many candidate phases have been discussed, such as various color superconductors \cite{Alford:2007xm}, suggested by perturbative results -- which are applicable at asymptotically large densities --  and by phenomenological models. Another candidate phase is quarkyonic matter, suggested by results for 
QCD at a large number of colors $N_c$ \cite{McLerran:2007qj,Andronic:2009gj,Fukushima:2013rx,Philipsen:2019qqm}. In this paper we investigate quarkyonic matter in a holographic top-down approach (within a certain approximation). In particular, we determine the phase structure of the model fully dynamically for all temperatures and baryon chemical potentials, i.e., we compare the free energies of quarkyonic matter with those of pure baryonic matter, pure quark matter, and the mesonic phase. 

There are various ways of characterizing and defining quarkyonic matter, and different studies in the literature focus, sometimes confusingly, on different, often hypothetical, properties of this phase. Originally \cite{McLerran:2007qj}, it was pointed out that the pressure of "large-$N_c$ nuclear matter" scales linearly with $N_c$, suggesting a bulk contribution of quarks, while confinement indicates that the fermionic excitations of the system are color singlets. Hence the term quarkyonic, suggesting that the phase is partly quark-like, partly baryonic. It was speculated that quarkyonic matter may be chirally symmetric, at least at sufficiently large densities. Therefore, several model calculations identify confined, but chirally symmetric matter with the quarkyonic phase, even when baryonic degrees of freedom are not included, such as in the Polyakov-Nambu-Jona-Lasinio model \cite{Fukushima:2008wg,McLerran:2008ua,Sakai:2011fa}. It was also argued that quarkyonic matter is spatially inhomogeneous due to the appearance of chiral density waves \cite{Kojo:2009ha,Kojo:2011cn}. Recently, 
a simple model of (chirally broken) quarkyonic matter was proposed \cite{McLerran:2018hbz}, based on the picture of a Fermi sea filled by quarks surrounded by a baryon layer. This model has been extended and applied to the physics of neutron stars \cite{Jeong:2019lhv,Sen:2020peq,Duarte:2020xsp,Zhao:2020dvu}.   

In our holographic approach, we propose a very simple construction of the quarkyonic phase, similar in spirit to the model of Ref.\ \cite{McLerran:2018hbz}: we find a geometric configuration, uniform in position space, in which quark and baryon degrees of freedom coexist, such that baryon density is generated by actual baryons {\it and} by fundamental quarks. Instead of a construction in momentum space, our quarkyonic matter is constructed in the bulk of the holographic model, with  quarks and baryons affecting different domains of the holographic direction, associated with different energy regimes. Here, baryons are restricted to the ultraviolet, while quarks dominate the infrared. Since we work at strong coupling, an exact reproduction of the Fermi surface picture realized in the weak-coupling model of Ref.\ \cite{McLerran:2018hbz} is not expected. Besides this realization of a quarkyonic phase in the literal sense of the word, our model also allows us to check dynamically whether quarkyonic matter is chirally symmetric and in which regions of the phase diagram it is the preferred configuration.

We employ the Witten-Sakai-Sugimoto model \cite{Witten:1998zw,Sakai:2004cn,Sakai:2005yt}, which is based on type-IIA string theory, and we work in the so-called deconfined geometry and the decompactified limit of the model. Baryons can be introduced as instanton configurations of the gauge theory on the flavor branes in the bulk. For simplicity, we approximate these instantons by delta peaks \cite{Bergman:2007wp}. Our main point is that these pointlike baryons, whose location on the branes is dynamically determined, can exist in geometries which already have a nonzero quark number density. Here, the quark density is either created by string sources attached to the tip of the connected flavor branes \cite{Bergman:2007wp,Kovensky:2019bih} or by a nontrivial abelian gauge field on the flavor branes reaching all the way to the horizon \cite{Bergman:2007wp,Aharony:2006da,Horigome:2006xu,Kovensky:2019bih}. It is known how to go beyond the pointlike approximation of baryonic matter, by allowing for nonzero instanton widths \cite{Ghoroku:2012am,Li:2015uea,Preis:2016fsp}, and including instanton interactions \cite{BitaghsirFadafan:2018uzs}. Our construction of the quarkyonic phase can be improved in the future by implementing these features. 

There exist previous studies regarding a holographic realization of the quarkyonic phase. Within the  Witten-Sakai-Sugimoto model it has been argued that quarkyonic matter can be described within a purely baryonic approach, as soon as the baryons move towards the holographic boundary  \cite{Kaplunovsky:2012gb,Kaplunovsky:2013iza}, and instabilities towards the creation of a quarkyonic phase have been looked for at parametrically large baryon densities 
\cite{deBoer:2012ij}. Our approach goes beyond these studies by including quark degrees of freedom and by constructing the quarkyonic phase explicitly. The quarkyonic phase has also been studied recently in a bottom-up holographic model 
\cite{Chen:2019rez}, where it is  identified with a chirally symmetric but confined phase, albeit without including baryons. In Ref.\ \cite{Ishii:2019gta}, another holographic bottom-up approach, working in the so-called Veneziano limit, was employed to construct a phase where quarks and pointlike baryons coexist. It was argued, however, that this phase only exists in a version of the model that is unfavored otherwise.

Some of our results are presented in the chiral limit, i.e., for a vanishing current quark mass. This limit is sufficient to demonstrate the existence of the holographic quarkyonic phase, and it can be used to derive some analytical results. Moreover, the case of an exact chiral symmetry is useful to identify and understand the chiral phase transition. We generalize these results by allowing for a nonzero current quark mass, resulting in a nonzero pion mass. Here we follow our recent work \cite{Kovensky:2019bih}, in which the effect of a quark mass  on the phase structure of the Witten-Sakai-Sugimoto model was explored -- however in the absence of any baryonic degrees of freedom. Massive quarks are of particular interest in the context of a possible quark-hadron continuity: in Ref.\ \cite{BitaghsirFadafan:2018uzs} it was speculated that the model allows for a continuous transition between nuclear and quark matter at zero temperature, which is conceivable in QCD \cite{Schafer:1998ef,Hatsuda:2006ps,Schmitt:2010pf,Baym:2019iky}, although usually not predicted by phenomenological models. Here we do find a realization of this continuity via the detour of quarkyonic matter: at sufficiently large values of the pion mass, quarks start to appear continuously at a certain chemical potential, while heating up the quarkyonic phase makes baryons melt, such that a pure quark phase is reached continuously. 

Our paper is organized as follows. After a brief summary of the main ingredients of the model in Sec.\ \ref{sec:setup}, we discuss the action in Sec.\ \ref{sec:action} and the resulting equations of motion in Sec.\ \ref{sec:EOMs}. Sec.\ \ref{sec:phases} introduces the possible candidate phases within our approach, and in Sec.\ \ref{sec:chiral} we prove analytically that holographic quarkyonic matter is energetically preferred over quark matter in the chiral limit and at zero temperature. In Sec.\ \ref{sec:thermo} we present the numerical evaluation, most notably the phase structure in Sec.\ \ref{sec:phasediagram}, while in Secs.\ \ref{sec:frac}, \ref{sec:embed}, and \ref{sec:continuity} we go into some details, including discussions of the brane embeddings, the quark-hadron continuity, and the speed of sound.

\section{Setup, candidate phases, and analytical results}

\subsection{Holographic model}
\label{sec:setup}

In the Witten-Sakai-Sugimoto model, the pure glue physics of the field theory is described in terms of the dual gravitational background. This background is given in terms of the metric with curvature radius  $R$, the dilaton and the Ramond-Ramond 4-form. It is sourced by $N_c$ D4-branes in type-IIA superstring theory. One of the extra dimensions, say $X_4$, is compactified on a circle of radius $R_4$ and the theory is only effectively four-dimensional for energies below the Kaluza-Klein scale $\MKK = R_4^{-1}$. Moreover, the periodicity conditions on the $X_4$-circle are chosen such that supersymmetry is broken. The resulting low-energy theory is believed to be in the same universality class as large-$N_c$ pure Yang-Mills theory. This background geometry undergoes a Hawking-Page transition as the temperature $T$ is increased. The corresponding critical temperature $T_c = \MKK/(2\pi)$ is usually identified with the deconfinement temperature. 

Fundamental matter is included by adding $N_f$ pairs of D8- and \textoverline{D8}-branes ("flavor branes"), such that D4-D8 and D4-\textoverline{D8} strings are associated with left- and right-handed fundamental fermions. At the lowest non-trivial order in $N_f/N_c$ these extra D-branes can be taken as probes in a fixed background. We shall work within this probe brane approximation throughout the paper. 
Asymptotically, i.e., for large values of the radial (holographic) direction $U$, branes and anti-branes are separated along the $X_4$ direction by a fixed distance $L$. This separation becomes dynamical in the bulk, and chiral symmetry breaking  occurs when branes and anti-branes join at some $U=U_c$, providing a geometrical realization of the spontaneous symmetry breaking pattern $U(N_f)_L \times U(N_f)_R \to U (N_f)$. A quark chemical potential $\mu_q$ is introduced by turning on an asymptotic value for the temporal component of the abelian gauge field on the D8-branes, $\hat{A}_0(\infty) = \mu_q$. For large chemical potentials, the results of the probe brane approximation have to be interpreted with caution since effects on the background geometry can become sizable.  

The free parameters of the model are $\MKK$, $L$, and the (four-dimensional) 't Hooft coupling $\lambda$. The gravitational description is most accurate at $\lambda\gg 1$. In the original version of the model \cite{Sakai:2004cn,Sakai:2005yt},
the flavor branes were chosen to be located at antipodal points of the $X_4$-circle, i.e., the asymptotic separation was fixed at $L=\pi M_{\rm KK}^{-1}$. In this case, deconfinement and chiral restoration are locked together. In a more general setup, if the flavor branes are non-antipodal, for $T>T_c$ (i.e., in the "deconfined geometry") it becomes a dynamical question whether the flavor branes  join in the bulk or not. For sufficiently small values of $L$ there is always a region in the $T$-$\mu_q$ plane where a deconfined but chirally broken phase is favored \cite{Aharony:2006da,Horigome:2006xu}. Moreover, the critical temperature for (approximate) chiral symmetry restoration depends on the chemical potential, as expected in real-world QCD.  
In this paper, following Refs.\ \cite{Preis:2010cq,Preis:2012fh,Preis:2016fsp,BitaghsirFadafan:2018uzs,Kovensky:2019bih}, we assume the background to be in the deconfined geometry for arbitrarily low $T$, which requires a small separation, $L\ll \pi M_{\rm KK}^{-1}$, and is referred to as the "decompactified limit". (Apart from restricting ourselves to the deconfined geometry, $L\ll \pi M_{\rm KK}^{-1}$ is never explicitly needed in our calculation.)  In this parameter regime, the gluons are effectively decoupled from the dynamics of chiral symmetry breaking. Therefore, the field-theoretical dual is comparable to a Nambu-Jona-Lasinio model \cite{Antonyan:2006vw,Davis:2007ka}. While it is true that by going to this regime we lose some control of the original top-down construction (in particular, the Kaluza-Klein modes are potentially relevant), we gain a much richer phase structure, likely to be closer to nature, at least with respect to the chiral phase transition.

\subsection{Building blocks of the action}
\label{sec:action}

Having stated the main ideas behind the model (for more details see the original works or the review \cite{Rebhan:2014rxa}), we now introduce the action. Since each term in the action has been considered before in the literature, we will not elaborate on any derivations. Moreover, from now on we work with dimensionless quantities, following the convention of Ref.\ \cite{Li:2015uea}: we use the coordinates $u=U/(R^3M_{\rm KK}^2)$, $x_4=X_4 M_{\rm KK}$, the abelian gauge field $\hat{a}_0=4\pi \hat{A}_0/(\lambda M_{\rm KK})$, and the asymptotic separation of the flavor branes $\ell= LM_{\rm KK}$. As a consequence, the thermodynamic quantities such as quark chemical potential $\mu$, baryon density $n$, and temperature $t$ will also be used in their dimensionless version. For some numerical estimates we will translate back to the dimensionful counterparts. This can be done with the help of Table 1 of Ref.\ \cite{Kovensky:2019bih}, see also Table 1 of Ref.\ \cite{Li:2015uea}.

We consider the action
\begin{equation}
    S = S_{\mathrm{DBI}} + S_{q} + S_b
    + S_{m} \, , 
    \label{fullS}
\end{equation}
with the Dirac-Born-Infeld (DBI) term $S_{\mathrm{DBI}}$, the contribution of string sources $S_q$, the contribution of pointlike baryons $S_b$, and the quark mass contribution $S_m$. Their explicit forms are as follows.

The DBI action  for the world-volume fields $x_4(u)$ and $\hat{a}_0(u)$ is  
    \begin{equation}
        S_{\mathrm{DBI}} = {\cal{N}}N_f \frac{V}{T} \int_{u_c}^{\infty} 
        du \, {\cal L}_{\rm DBI} \, ,\label{DBI}
    \end{equation}
    where $V$ is the 3-volume of the flat spatial directions, and 
    \begin{equation} \label{N}
        {\cal{N}} \equiv \frac{N_c \MKK^4\lambda_0^3}{6\pi^2} \,, 
    \end{equation}
    with $\lambda_0 = \lambda/(4\pi)$. We have introduced the DBI Lagrangian
    \be \label{LDBI}
    {\cal L}_{\rm DBI} = u^{5/2} \sqrt{1 + u^3 f_T(u) x_4'^2(u) - \hat{a}_0'^2(u)} \, , 
    \ee
    where primes denote derivatives with respect to $u$, and $f_T(u)$ is the blackening factor of the background metric,
    \begin{equation}
        f_T(u) = 1 - \frac{u_T^3}{u^3} \, .   
    \end{equation}
    Here, $u_T$ denotes the location of the horizon of the background geometry and is related to the dimensionless temperature $t$ by  
    \begin{equation} \label{tuT}
      t = \frac{3}{4\pi} \sqrt{u_T} \, .   
    \end{equation}
   The DBI action is written for the case of flavor branes joining at $u=u_c$, where $u_c$ has to be determined dynamically. If the flavor branes reach all the way to the horizon, the lower boundary of the integral has to be replaced by $u_T$, which is fixed by the temperature. In the chiral limit, these two different geometries correspond to chirally broken and chirally restored phases, respectively. We stick to the formulation of joined branes at $u_c$ until we need to discuss both cases separately in Sec.\ \ref{sec:phases}. In either case, the prefactor ${\cal N}$  includes a factor of $2$ accounting for the two halves of the configuration. We shall write all terms of the action such that ${\cal{N}} N_f$ becomes a global prefactor. In particular, we only consider flavor-symmetric systems, i.e., even when we consider nonzero quark masses we assume them to be degenerate for all flavors.

\begin{figure}[t]
    \centering
    \includegraphics[width=0.7\textwidth]{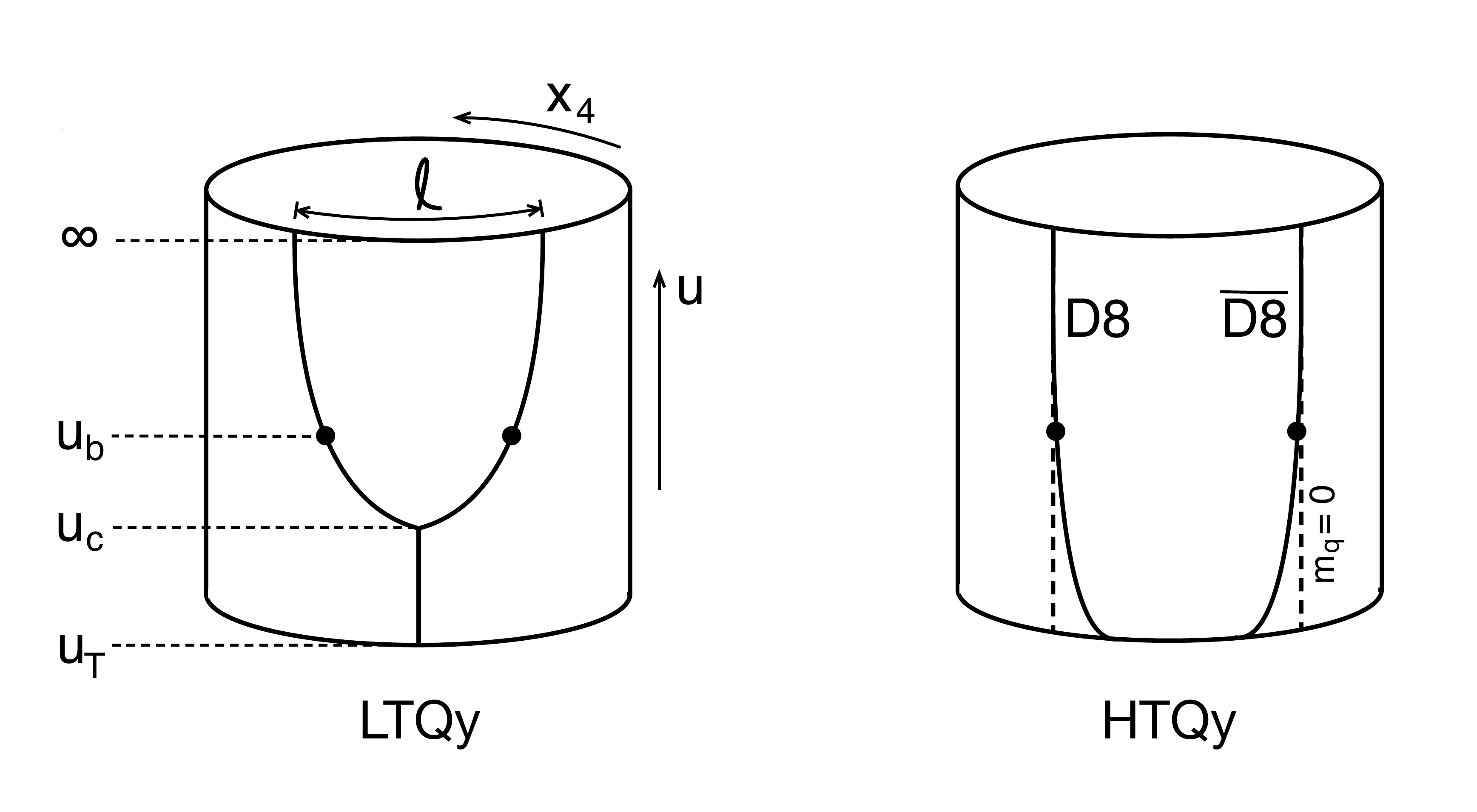}
    \caption{{\small Geometric configurations of holographic quarkyonic matter in the $x_4$-$u$ subspace. In the low-temperature quarkyonic phase (LTQy), strings stretch from $u_c$ to the horizon at $u_T$, while baryons (assumed to be pointlike) are located at $u_b$. In the high-temperature 
    quarkyonic phase (HTQy), the flavor branes extend down to the horizon (they are straight in the absence of a current quark mass $m_q$), and again baryons sit at some $u_b$. The
    locations $u_c$ and $u_b$ and the embedding of the D8- and \textoverline{D8}-branes (which have cusps at $u_b$ and $u_c$) are determined dynamically for given temperature and chemical potential. Other candidate phases -- previously discussed in the literature -- are obtained by removing string sources and/or baryons. }} 
    \label{fig:quarkyonic}
\end{figure}

The contribution $S_q$ accounts for $N_s$ strings stretching vertically from the horizon at $u_T$ to the tip of the embedding of the D8-\textoverline{D8} pairs at $u_c$, see left panel of Fig.\ \ref{fig:quarkyonic}. They represent a source for the gauge field $\hat{a}_0$ and give rise to a nonzero baryon number density $n_q$. Since the strings carry color and flavor charges like fundamental quarks we have added the subscript $q$ and we will interpret $n_q$ as baryon number created by quarks. This does not necessarily suggest a quasiparticle picture in the dual field theory, which, due to the strong coupling limit, cannot be expected to be valid.
We restrict ourselves to non-negative chemical potentials, and thus assume all contributions to the baryon density to be non-negative as well. In principle, one might allow for the strings to attach to the flavor branes at more than a single point. And in fact, for the baryonic sources, we shall exactly do so. We have checked, however, that a configuration with two sets of strings, attached symmetrically to the two halves of the flavor branes is never favored energetically. Therefore, we continue with strings attached only at $u=u_c$. Also assuming the density to be uniform in the flat directions, the Nambu-Goto term and the boundary contribution yield \cite{Bergman:2007wp,Kovensky:2019bih}
    \begin{equation}
    S_{q} = {\cal{N}}N_f \frac{V}{T} \int_{u_c}^{\infty} du\, n_q \left[
    (u - u_T) - \hat{a}_0(u)\right]\delta(u-u_c) \, ,
    \label{Sstring}
    \end{equation}
where we have defined the dimensionless baryon number density (per flavor) created by string sources by
    \begin{equation}
    n_q = \frac{6\pi^2}{\lambda_0^2M_{\rm KK}^3N_fN_c}\frac{N_s}{V} \, .   
    \end{equation}
Here, $N_s/N_c$ is the baryon number from strings, and the prefactor arises because we have pulled out a factor ${\cal N}N_f$. The energy of the strings is given by the Nambu-Goto term in Eq.\ (\ref{Sstring}), and thus we can define the dimensionless constituent quark mass by \cite{Aharony:2008an,Argyres:2008sw,Kovensky:2019bih}
\be \label{quarkmass}
M_q = u_c-u_T \, .
\ee
This mass is medium dependent, not only through the explicit temperature dependence, but also through $u_c$, which implicitly depends on temperature and chemical potential.

If the flavor branes reach $u=u_T$, we need to impose the boundary condition $\hat{a}_0(u_T)=0$, and $S_q$ plays no role. Nevertheless, although the string sources are absent, this case also allows for a nonzero quark density due to the nontrivial profile of $\hat{a}_0$. This scenario is shown in the right panel of Fig.\ \ref{fig:quarkyonic}.

Next, $S_b$ is the contribution from baryons. In the holographic description, baryons are effectively described in terms of D4-branes wrapping the four-sphere $S^4$ in the ten-dimensional geometry \cite{Sakai:2004cn,Hata:2007mb} (not to be confused with the $N_c$ D4-branes of the background geometry). 
As there are $N_c$ units of 4-form flux on this $S^4$, the same amount of fundamental strings are attached to the D4-branes. These strings can end on 
the flavor branes, and due to their strong tension their length is minimized, forcing the baryon to live on the D8-\textoverline{D8} embedding \cite{Seki:2008mu}. From the point of view of the world-volume theory, they can be viewed as non-abelian instantons in the four-dimensional space spanned by $\vec{x}$ and $u$, and baryon charge is identified with the topological instanton number \cite{Hata:2007mb}. This is manifest in the Chern-Simons (CS) term of the D8-brane action, which couples $\hat{a}_0$ with the topologically non-trivial configuration of the non-abelian gauge fields. Here we work with the simplest approximation, assuming the instantons to be pointlike and non-interacting \cite{Bergman:2007wp} (which does not imply that the baryons in the dual field theory are non-interacting). The action receives contributions from the energy of the D4-branes and from the Chern-Simons term. Again restricting ourselves to uniform distributions in the flat direction and assuming the baryon density to be non-negative, this results in \cite{Bergman:2007wp,Preis:2011sp,Li:2015uea}
\begin{equation} \label{Sb}
    S_b = S_{\mathrm{D4}} + S_{\mathrm{CS}} = {\cal{\N}} N_f \frac{V}{T}  \int_{u_c}^{\infty} du\, n_b  \left[
    \frac{u}{3}\sqrt{f_T(u)} - \hat{a}_0(u)\right]\delta(u-u_b) \, . 
\end{equation}
We have denoted the location of the baryons in the holographic direction by $u_b$, and the number of pointlike instantons $N_I$ is related to the dimensionless baryon density by 
\begin{equation}
    n_b = \frac{6 \pi^2}{\lambda_0^2 \MKK^3 N_f} \frac{N_I}{V} \, .
\end{equation}
Since the first term in Eq.\ (\ref{Sb}) corresponds to the energy of the D4-branes, we can define the dimensionless baryon mass by
\be \label{baryonmass}
M_b = \frac{u_b}{3}\sqrt{f_T(u_b)} \, .
\ee
Just like the constituent quark mass (\ref{quarkmass}), this mass varies with chemical potential and temperature, i.e., it is the mass of the baryon in the medium.

 Originally \cite{Bergman:2007wp}, the instantons were assumed to be located at the tip of the connected flavor branes, $u_b=u_c$. However, it has been realized that one has to allow for the instantons to spread out in the holographic directions, i.e., more than one baryon layer may appear \cite{Kaplunovsky:2012gb,Kaplunovsky:2013iza,Elliot-Ripley:2016uwb,Preis:2016fsp}. In general, each layer has a location in the holographic direction $u_b$ and a contribution to the baryon density $n_b$ and all locations and densities have to be determined dynamically. Here, we restrict ourselves to the 1-layer case, $u_b=u_c$, and the 2-layer case, where a single $u_b>u_c$ is determined dynamically (there is one layer on each half of the flavor branes at the same $u_b$). We shall see that the 1-layer solution is never the state of lowest free energy.  The 2-layer solution, however, is crucial for our purposes: it allows us to combine baryons from instantons with quarks from string sources (if the flavor branes connect at $u_c$) and also to place baryons into the geometry if the D8- and $\overline{\rm D8}$-branes reach the horizon, giving rise to two geometric realizations of quarkyonic matter, see Fig.\ \ref{fig:quarkyonic}. 
 
In principle, we would have to allow for a larger number of layers, which obviously becomes tedious if they are treated in full generality. We have checked that higher numbers of layers do become favored in the present approximation, as pointed out in Ref.\ \cite{Kaplunovsky:2012gb}. However, our main conclusions (e.g., existence of and transition to the quarkyonic phase) remain unaltered. Furthermore, it is not known if a higher number of layers is indeed favored in less simplistic approximations. For instance, two approximations different from Ref.\ \cite{Kaplunovsky:2012gb} and different from each other  \cite{Elliot-Ripley:2016uwb,Preis:2016fsp} indicate that more than 2 layers are never preferred. From now on we shall therefore ignore configurations with 3 or more instanton layers.

The final term in Eq.\ (\ref{fullS}), $S_{m}$, accounts for a (small) current quark mass $m_q$ and is taken from Ref.\ \cite{Kovensky:2019bih}, where more details about its derivation can be found. Introducing a current quark mass  in the Witten-Sakai-Sugimoto model is not straightforward (as opposed to a constituent quark mass, which is dynamically generated by chiral symmetry breaking). This is mainly due to the fact that there are no extra directions along which color and flavor branes can be separated, as the usual geometrical version of the Higgs mechanism would require. Here we employ the approximation based on a gauge-invariant, albeit non-local, version of the quark bi-linear operator whose expectation value is related to the chiral condensate $\langle \bar{q} q\rangle$ \cite{Aharony:2008an,Hashimoto:2008sr,McNees:2008km}. This approach reproduces the pion mass term from chiral perturbation theory in the low energy description, as well as the Gell-Mann-Oakes-Renner relation. At strong coupling and from the holographic point of view, the expectation value of this operator is computed in terms of the area swept by a string (more precisely, a world-sheet instanton) stretched between the D8- and the \textoverline{D8}-branes. The mass contribution to the action has the form $S_m\sim m_q\langle \bar{q} q\rangle$, and, employing the notation of Ref.\ \cite{Kovensky:2019bih}, it can be written as 
\begin{equation} \label{S_m}
S_{m} = -{\cal N}N_f\frac{V}{T}\frac{A}{2\lambda_0} \, ,
\end{equation}
where 
\begin{equation}
A\equiv \frac{2 \al}{\lambda_0^2} \, \exp\left\{2\lambda_0\left[\phi_T(u_c)x_4(u_c)  + \int_{u_c}^{\infty} du \, \phi_T(u) x_4'(u) \right]\right\} \, .
\label{Adef1}
\end{equation}
The exponent corresponds to the Nambu-Goto action of the string, which has already been renormalized by subtracting the vacuum contribution, as explained in detail in Ref.\ \cite{Kovensky:2019bih}. 
The first term in the exponential vanishes if the flavor branes connect at $u_c$, since $x_4(u_c)=0$, but it becomes non-trivial if they reach the horizon, in which case $x_4(u_T)\ge 0$. We have abbreviated  
\begin{equation}
    \phi_T(u) \equiv \int \frac{du}{\sqrt{f_T(u)}} = \frac{u}{\sqrt{f_T(u)}} \left\{
    1-\frac{3u_T^3}{4 u^3 f_T^{1/6}(u)} {}_2F_1 \left[
    \frac{1}{6},\frac{2}{3},\frac{5}{3}, -\frac{u_T^3}{u^3 f_T(u)}
    \right]\right\} \, , 
\end{equation}
where ${}_2F_1$ is the hypergeometric function, and the dimensionless "mass parameter" $\alpha$ is
\begin{equation}
\alpha = \frac{3 \pi^2 f_\pi^2 m_\pi^2}{N_c \MKK^4}\exp\left(-\frac{\lambda_0}{\ell}\pi\tan\frac{\pi}{16}\right) \, , \label{defalpha} 
\end{equation} 
with the pion mass $m_\pi$ and the pion decay constant $f_\pi$. 
In order to translate the value of $\alpha$ into a value of the pion mass it is useful to compute the pion decay constant within our model. This is done by deriving an effective action for the pions from the Yang-Mills approximation of the DBI action and identifying the prefactor of the kinetic term with the pion decay constant squared (divided by 4). Following Refs.\ \cite{Sakai:2004cn,Callebaut:2011ab}, we find for the deconfined geometry 
\be
f_\pi^2 = \frac{\lambda N_c M_{\rm KK}^2}{48\pi^3} \left(\int_{u_c}^\infty du\,  \frac{\sqrt{1+u^3x_4'^2}}{u^{5/2}}\right)^{-1} \, ,
\ee
where $x_4(u)$ is the embedding function of the flavor branes in the vacuum, i.e., at zero temperature and chemical potential. Employing the massless limit, we have (see for instance Appendix B of Ref.\ \cite{Li:2015uea}) 
\be
x_4'^2(u)=\frac{u_{c0}^8}{u^3(u^8-u_{c0}^8)} \, , \qquad u_{c0} = \frac{16\pi}{\ell^2}\left(\frac{\Gamma[9/16]}{\Gamma[1/16]}\right)^2 \, ,
\ee
which allows us to evaluate the integral analytically to obtain
\be
f_\pi^2 = \frac{32 \lambda N_c M_{\rm KK}^2}{3\pi^2\ell^3} \left(\frac{\Gamma[9/16]}{\Gamma[1/16]}\right)^3\frac{\Gamma[11/16]}{\Gamma[3/16]}\, . \label{fpi}
\ee

The quark mass term (\ref{S_m}) receives an additional contribution if there is a nonzero radial component of the  world-volume gauge field $a_u$. In principle, such a contribution is relevant in our context because $a_u$ is nonzero for instance in the usual Belavin-Polyakov-Schwarz-Tyupkin instanton configuration \cite{1975PhLB...59...85B}.  This correction was used to compute the baryon mass shift and the spectrum modifications induced by $m_q$ \cite{Hashimoto:2009hj}. However, as can be seen in Eq.~(2.18) of this reference, the correction is proportional to the cube of the instanton width. Since we work with pointlike instantons throughout the paper, this term vanishes in our approximation, and we may proceed with Eq.\ (\ref{S_m}).

\subsection{Equations of motion and free energy} 
\label{sec:EOMs}

Inserting the expressions (\ref{DBI}), (\ref{Sstring}), (\ref{Sb}), and (\ref{S_m}) into Eq.\ (\ref{fullS}) gives the 
full action. The equations of motion 
\be
\frac{\delta S}{\delta x_4(u)} = \frac{\delta S}{\delta \hat{a}_0(u)} = 0 
\ee
become
\begin{subequations}\label{DelPi}
\bea
0&=& \partial_u\pi_{x_4}+\Delta\pi_{x_4}\delta(u-u_b) \, ,  \\[2ex]
0&=& \partial_u\pi_{\hat{a}_0}+(\Delta\pi_{\hat{a}_0}+n_b)\delta(u-u_b) +n_q\delta(u-u_c) \, ,
\eea
\end{subequations}
where the canonical momenta,
\be
\pi_{x_4}(u) \equiv \frac{\partial {\cal L}_{\rm DBI}}{\partial x_4'}-A\phi_T(u)  \, , \qquad \pi_{\hat{a}_0}(u) \equiv \frac{\partial {\cal L}_{\rm DBI}}{\partial \hat{a}_0'} \,, 
\ee
have been allowed to be discontinuous due to the baryon source at $u=u_b$, and we have abbreviated $\Delta\pi \equiv \pi(u_b^+)-\pi(u_b^-)$. 
Integrating the equations of motion over $[u_c,u]$ yields 
\begin{subequations} \label{EOM}
\bea
  \pi_{x_4}(u) &=&  \frac{u^{11/2} f_T(u) x_4'(u)}{\sqrt{1+u^3 f_T(u)x_4'^2(u) - \hat{a}_0'^2(u)}} -A \phi_T(u) =  k \, , \label{EOM1} \\[2ex]
  -\pi_{\hat{a}_0}(u)&=&  \frac{u^{5/2} \hat{a}_0'(u)}{\sqrt{1+u^3 f_T(u)x_4'^2(u) - \hat{a}_0'^2(u)}} = n_q + n_b \Theta (u - u_b) \, , \label{EOM2}
\eea
\end{subequations}
where $k=\pi_{x_4}(u_c)$ is an integration constant and $\Theta$ is the Heaviside step function. The discontinuities in Eqs. (\ref{DelPi}) have canceled since they are exactly reproduced with opposite sign by the $u$-integration over $\partial_u\pi_{x_4}$ and $\partial_u\pi_{\hat{a}_0}$. As a consequence, there is only a single integration constant in Eq.\ (\ref{EOM1}), not two different ones depending on whether $u<u_b$ or $u>u_b$. In Eq.\ (\ref{EOM2}), the baryon density from baryons $n_b$ originates from the pointlike instantons at $u=u_b$, with the step function appearing due to the integral over $\delta(u-u_b)$. As a consequence, $x_4(u)$ and $\hat{a}_0(u)$ do not depend on $n_b$ for $u<u_b$. The baryon density from quarks $n_q$ arises in the presence of string sources due to the integral over $\delta(u-u_c)$. In this case, no additional integration constant appears. This is best seen by repeating the calculation with string sources at some $u_q>u_c$ and at the end letting $u_q\to u_c$. Alternatively, if string sources are absent and the flavor branes reach the horizon at $u_T$, there is an additional integration constant which is identified with the baryon density from quarks, $n_q = -\pi_{\hat{a}_0}(u_T)$, such that we can use Eq.\ (\ref{EOM2}) for both cases.

We can solve Eqs.\ (\ref{EOM}) algebraically for $x_4'(u)$ and $\hat{a}_0'(u)$,
\begin{subequations} \label{EOMs}
\bea
  x_4'(u) &=& \frac{A \phi_T(u) + k}{u^{11/2} f_T(u)} \zeta(u)\, , 
  \label{solx4}\\[2ex] 
  \hat{a}_0'(u) &=& \frac{n_q + n_b \Theta(u - u_b)}{u^{5/2}} \zeta(u)\, ,  \label{sola0}
 \eea
\end{subequations}
where we have abbreviated 
\bea \label{zeta}
    \zeta(u) &\equiv&\sqrt{1 + u^3 f_T(u) x_4'^2(u) - \hat{a}_0'^2(u)} \non[2ex]
    &=& \left\{1-\frac{[A \phi_T(u) + k]^2}{u^8 f_T(u)} + \frac{\left[
    n_q + n_b \Theta(u - u_b)
    \right]^2}{u^5}\right\}^{-1/2}   
    \, .
\eea
The boundary conditions at the holographic boundary, $x_4(\infty)=\ell/2$ and $\hat{a}_0(\infty)=\mu$, are written most conveniently in the form
\begin{subequations}\label{bc}
\bea
    \label{bc1}
    \frac{\ell}{2} &=& \int_{u_c}^{\infty} du \, x_4'(u) + x_4(u_c) \, , \\[2ex]
    \mu 
   &=&\int_{u_c}^{\infty}du \, \hat{a}_0'(u) + \hat{a}_0(u_c) = \int_{u_b}^{\infty}du \, \hat{a}_0'(u) + \hat{a}_0(u_b)
     \, .\label{bc2}
\eea
\end{subequations}
These relations are only valid if $x_4(u)$ and $\hat{a}_0(u)$ are continuous at $u=u_b$. We shall require continuity in all cases we consider. However, the derivatives $x_4'(u)$ and $\hat{a}_0'(u)$ are
necessarily discontinuous in the presence of pointlike charges, inducing cusps in the embedding of the flavor branes. If the flavor branes connect as in the left panel of Fig.\ \ref{fig:quarkyonic}, $x_4(u_c)=0$, and $\hat{a}_0(u_c)$ is nonzero and needs to be determined dynamically. If, on the other hand, the branes reach the horizon, as in the right panel of Fig.\ \ref{fig:quarkyonic}, $u_c$ has to be replaced by $u_T$, $x_4(u_T)\ge 0$ has to be determined dynamically, and we have to impose the additional boundary condition $\hat{a}_0(u_T)=0$.   
From the behavior of $\hat{a}_0(u)$ close to the holographic boundary one reads off the density associated with the chemical potential $\mu$, via the usual AdS/CFT dictionary. Indeed, by expanding $\hat{a}_0'(u)$ from Eq.\ (\ref{sola0}) for large $u$ we find that the total baryon density is 
\be
n\equiv n_q+n_b \, .
\ee
It is useful to introduce the baryon fraction 
\be \label{xbdef}
x_b\equiv \frac{n_b}{n} \, , 
\ee
such that $0\le x_b\le 1$. In the numerical evaluation it turns out to be somewhat more convenient to work with $x_b$ and $n$ rather than with $n_b$ and $n_q$. 

The free energy of the system is given by the on-shell action.  By re-inserting the solutions (\ref{EOMs}) into the action 
(\ref{fullS}), adding and subtracting $k [\ell/2 - x_4(u_c)] -\mu n $, and rewriting $\ell$ and $\mu$ with the help of Eqs.\ (\ref{bc}), we obtain the dimensionless free energy density in the convenient form 
\begin{eqnarray}
   \Omega &\equiv& \frac{S\big|_{\rm on-shell}}{{\cal N}N_f \frac{V}{T}} = \int_{u_c}^{\infty}du\, \left[
 \frac{u^{5/2}}{ \zeta(u)} +
 A \phi_T(u) x_4'(u) \right] 
 - \frac{A}{2\lambda_0} \non[2ex]
&&  - \mu n + n_q (u_c - u_T) + n_b \frac{u_b}{3}\sqrt{f_T(u_b)}
 + k \left[\frac{\ell}{2} - x_4(u_c)\right] \,. \label{Omega} 
\end{eqnarray}
This expression is formally divergent, but it is straightforwardly  renormalized by subtracting the vacuum contribution, i.e., we regularize the integral by introducing an ultraviolet cutoff $\Lambda$, subtract the vacuum free energy density $\Omega_{t=\mu=0}=\frac{2}{7}\Lambda^{7/2}$ and then take the $\Lambda \to \infty$ limit.  (Recall that in our effective treatment of the mass correction the Nambu-Goto action in the mass term \eqref{Adef1} had to be renormalized separately.)

\subsection{Stationarity equations and candidate phases}
\label{sec:phases}

For fixed mass parameter $\alpha$, temperature $t$, and chemical potential $\mu$ (and fixed model parameters $\ell$, $M_{\rm KK}$, $\lambda$), the free energy density is a function of the variables
$k$, $u_c$, $u_b$, $n_q$, $n_b$ if the flavor branes join at 
$u=u_c$ and the variables $k$, $u_b$, $n_q$, $n_b$, $x_4(u_T)$
if the flavor branes reach the horizon. We will now treat these two cases separately. Recall that $A$ in Eq.\ (\ref{Omega}) is an implicit function of all dynamical variables. (In the chiral limit, $A=0$.) We start with the case of connected flavor branes at $u_c>u_T$, i.e., $x_4(u_c)=0$. Requiring stationarity of $\Omega$ with respect to all dynamical variables yields
\begin{subequations}\allowdisplaybreaks
\label{min}
\begin{eqnarray}
    0 &=& \frac{\der \Omega}{\der k} = \frac{\ell}{2} - \int_{u_c}^{\infty}du
    \, x_4'(u) \, ,
    \label{mink} \\[2ex] 
    0 &=& \frac{\der \Omega}{\der n_q} = u_c - u_T - \hat{a}_0(u_c)\, ,
    \label{minnq} \\ [2ex] 
    0 &=&\frac{\der \Omega}{\der n_b} = 
    \frac{u_b}{3}\sqrt{f_T(u_b)} - \hat{a}_0(u_b) \, , 
    \label{minnI}\\ [2ex] 
    0 &=& \frac{\der \Omega}{\der u_c} = n_q - u_c^{5/2} \sqrt{1-\frac{[A \phi_T(u_c) + k ]^2}{u_c^8 f_T(u_c)} + \frac{n_q^2}{u_c^5}} \, , 
    \label{minuc} \\[2ex] 
    0 &=&\frac{\der \Omega}{\der u_b} = \frac{n_b}{3} \Delta_T(u_b) 
   \non[2ex] 
    && + \, u_b^{5/2}\left\{ \sqrt{1-\frac{[A \phi_T(u_b) + k ]^2}{u_b^8 f_T(u_b)} + \frac{n_q^2}{u_b^5}} - \sqrt{1-\frac{[A \phi_T(u_b) + k ]^2}{u_b^8 f_T(u_b)} + \frac{\left(n_q + n_b\right)^2}{u_b^5}} \right\}  \, , \hspace{0.8cm} \label{minuI} 
\end{eqnarray}
\end{subequations}
where we have abbreviated 
\begin{equation}
    \Delta_T (u_b) \equiv \frac{\der }{\der u_b} \left[u_b \sqrt{f_T(u_b)}\right] = 
    \frac{1}{\sqrt{f_T(u_b)}}\left(
    1+ \frac{u_T^3}{2u_b^3}
    \right)
    \, .
\end{equation}
We see that stationarity with respect to $k$ \eqref{mink} is equivalent to the boundary condition for $x_4(u)$ \eqref{bc1}. The conditions \eqref{minnq} and \eqref{minnI} determine the value of the gauge field $\hat{a}_0 (u)$ at the location of the sources. As a consequence of these conditions, the source terms $S_q$ and $S_b$ in \eqref{fullS} do not contribute explicitly to the free energy, only implicitly  through the equations of motion. Therefore, at the stationary point, the free energy density can be computed from 
\be\label{Omstat1}
\Omega = \int_{u_c}^\infty du\, u^{5/2}\zeta(u) -\frac{A}{2\lambda_0} \, ,
\ee
with $\zeta(u)$ from Eq.\ (\ref{zeta}).
Stationarity with respect to $u_c$ (\ref{minuc}) and 
$u_b$ (\ref{minuI}) can be interpreted as force balance conditions in the $u$ direction at the points where the embedding of the flavor branes develop a cusp, as discussed for pointlike baryons in Ref.\ \cite{Bergman:2007wp}.
One can check that the force balance in the $x_4$ direction is trivially satisfied by construction. 

Some of these equations can be brought into a more convenient form: we use 
the boundary condition for $\hat{a}_0(u)$ \eqref{bc2} in Eqs.\ (\ref{minnq}) and (\ref{minnI}) and rewrite the resulting integral in Eq.\ (\ref{minnq}) with the help of Eq.\ (\ref{minnI}). Furthermore, we solve \eqref{minuc} for $k$, which gives a result independent of $n_q$ (using $n_q\ge 0$), and rewrite Eq.\ (\ref{minuI}) in terms of the total baryon density $n$ and the baryon fraction $x_b$, after which we can solve this equation for $n$. Thus we arrive at the following set of coupled equations, 
\begin{subequations} \allowdisplaybreaks \label{min1}
\begin{eqnarray}
    \frac{\ell}{2} &=& \int_{u_c}^{\infty}du
    \, x_4'(u) \, ,  \label{mink1} \\[2ex]
  \int_{u_c}^{u_b} du \, \hat{a}_0'(u) &=&   \frac{u_b}{3}\sqrt{f_T(u_b)} - u_c + u_T 
     \, ,\label{minnq1} \\[2ex]
   \int_{u_b}^{\infty} du \, \hat{a}_0'(u) &=&  \mu - \frac{u_b}{3}\sqrt{f_T(u_b)}   \, , \label{minnI1}\\[2ex]
k &=& u_c^4 \sqrt{f_T(u_c)} - A \phi_T(u_c) \, , \label{minuc1} \\[2ex]
n &=& \frac{6\Delta_T(u_b)}{u_b^{3/2}} \sqrt{\frac{
u_b^8 f_T(u_b) - [A \phi_T(u_b) + k]^2 }{
f_T(u_b) [9-\Delta_T^2(u_b)] [9 (2-x_b)^2-x_b^2\Delta_T^2(u_b)]
}} \, ,\label{minuI1} 
\end{eqnarray}
\end{subequations}
which have to be solved together with the implicit equation for $A$ (\ref{Adef1}),
\be
 A= \frac{2 \al}{\lambda_0^2} \, \exp\left[2\lambda_0 \int_{u_c}^{\infty} du \, \phi_T(u) x_4'(u) \right] \, . \label{eqA}
\ee
We can insert $k$ and $n$ from Eqs.\ (\ref{minuc1}) and (\ref{minuI1}) into the other equations to be left with (at most) 4 coupled equations to be solved numerically. The equations allow for several distinct phases, obtained by setting both, one, or none of the contributions $n_q$ and $n_b$ to the baryon density to zero. If $n_q$ and/or $n_b$ are zero, the corresponding stationarity equations, i.e., (\ref{minnq}) and/or (\ref{minnI}) should be ignored. The reason is that we have restricted ourselves to $n_q,n_b\ge 0$, and thus zero density corresponds to the boundary of our multi-variable space. Therefore, zero density can correspond to a minimum of the free energy even though the free energy behaves linearly as a function of the density at that point. (Had we allowed for negative densities -- as it was done for instance in Ref.\ \cite{Preis:2011sp} for pointlike baryons -- the free energy would be $\land$- or $\lor$-shaped at the zero-density point.) All phases except for the quarkyonic phase, where both $n_q$ and $n_b$ are nonzero, have been discussed previously in the literature, at least in the chiral limit $A=0$. For completeness, and to show that they can be obtained from our more general setup, we list all relevant phases, i.e., the phases that play a role in the phase diagram.

\begin{enumerate}[$(a)$]
    \item {\it Mesonic phase: $n_q = n_b = 0$.}
    
    In this case, the baryon density vanishes. As a consequence, the embedding 
    of the flavor branes $x_4(u)$ is smooth everywhere, $\hat{a}_0(u) = \mu$ is constant, and the free energy does not depend on $\mu$ and $u_b$. This phase is therefore evaluated by solving  Eqs.\ (\ref{mink1}), (\ref{minuc1}), and (\ref{eqA}) (while Eq.\ (\ref{minuI}) is automatically fulfilled for $n_b=0$). We can determine the point of a continuous onset of baryon density from the mesonic geometry by taking the zero-density limit of Eqs.\ (\ref{minnq}) and (\ref{minnI}): strings can be  inserted at the point
    \be
    n_q \;\;{\rm onset}: \qquad \mu=u_c-u_T \, , 
    \label{qonsetmes}
    \ee
    while pointlike instantons are inserted at $u_b=u_c$ and 
    \be \label{nbonset}
    n_b \;\;{\rm onset}: \qquad \mu=\frac{u_c}{3}\sqrt{f_T(u_c)} \, .
    \ee
    These conditions simply mean that quarks (baryons) can be added to the system if the chemical potential becomes equal or larger than the constituent quark mass (baryon mass), see Eqs.\ (\ref{quarkmass}) and (\ref{baryonmass}). In both conditions $u_c$ depends implicitly on the temperature, and no simple analytic expression in terms of temperature only can be obtained in general. It should also be emphasized that these chemical potentials do not necessarily correspond to a physical phase transition because the continuous change of geometry can occur in a metastable regime, i.e., there is another phase, or even another branch of the same phase, which has lower free energy.

     \item {\it Baryonic phase (with 2 layers): $n_q = 0$, $n_b > 0$.}
     
     Here, strings are absent, and pointlike instantons are located at $u_b\ge u_c$. The embedding is smooth at $u_c$ (unless $u_b=u_c$) but has a cusp at $u_b$. The relevant equations to be solved simultaneously are Eqs.\ (\ref{mink1}), (\ref{minnI1}),  (\ref{minuc1}), (\ref{minuI1}), and (\ref{eqA}). 
     As we see from Eq.\ (\ref{sola0}), for $n_q=0$ the gauge field $\hat{a}_0(u)$ is constant for all $u<u_b$. Using this observation together with Eq.\ (\ref{minnq1}), we see that quark number can be continuously added at the point given implicitly by 
     \be \label{nqonset}
    n_q \;\;{\rm onset}: \qquad \frac{u_b}{3}\sqrt{f_T(u_b)}=u_c-u_T \, .
    \ee
    For instance, at zero temperature, where $u_T=0$ and $f_T(u)\equiv 1$, strings can be attached at the tip of the connected flavor branes when the baryons sit at $u_b=3u_c$. 
    Since the right-hand side of Eq.\ (\ref{nqonset}) is the constituent quark mass (\ref{quarkmass}), and the left-hand side is identical to $\hat{a}_0(u_b)$, we see that $\hat{a}_0(u_b)$ plays the role of an effective chemical potential for the quarks. We will come back to this observation in Sec.\ \ref{sec:embed}.
     
     \item {\it Low-temperature quark phase (LTQ): $n_q > 0$, $n_b = 0$.} 
     
     In this case, baryon number is only generated by quarks. This phase was introduced in Ref.\ \cite{Kovensky:2019bih} and termed LTQ phase because it turns out that it only exists at sufficiently small temperatures. It has a high-temperature counterpart to be discussed below. (In the chiral limit, the LTQ phase is never stable, only its counterpart is relevant and exists for all temperatures.) The equations to be solved  are Eqs.\ (\ref{mink1}), (\ref{minnq}),  (\ref{minuc1}), and (\ref{eqA}). (As written, Eq.\ (\ref{minnq1}) cannot be used since it was derived using the stationarity equation with respect to $n_b$, hence we have to go back to Eq.\ (\ref{minnq}), into which we may insert the boundary condition (\ref{bc2}).) Also from this phase it is possible to go continuously to the phase where both $n_q$ and $n_b$ are nonzero, i.e., in the geometry where strings are attached to the flavor branes one can insert pointlike baryons with infinitesimally small density at some $u_b>u_c$ (for which there is no simple analytical expression). This continuous transition is indeed realized for large current quark masses and certain nonzero temperatures. 
     
\item {\it Low-temperature quarkyonic phase (LTQy): $n_q,n_b > 0$.}

This phase is characterized by the presence of both instantons and strings, i.e., both $n_q$ and $n_b$ are non-zero. This phase is a combination of the baryonic and LTQ configurations, where embedding and location of the instantons are adjusted dynamically to accommodate both kinds of baryon densities. All equations (\ref{min1}) and (\ref{eqA}) need to be taken into account here. Due to the presence of quark and baryonic contributions to the thermodynamics of the system, this phase is \textit{quarkyonic} in the literal sense of the word.

\end{enumerate}

Besides the four phases listed here, there is also the 1-layer baryonic phase. In this case, the baryons are located at $u=u_c$. Since this is the point at which the strings are assumed to be attached in the above derivation, the relevant stationarity equations cannot be straightforwardly extracted from Eqs.\ (\ref{min}). However, by repeating the derivation without strings it is straightforward to derive the stationarity equations. This constitutes the generalization of Ref.\  \cite{Bergman:2007wp} to nonzero current quark masses. For all values of the current quark mass we consider, the 1-layer baryonic phase turns out to be never preferred in any region of the phase diagram. In particular, we find that the onset of 1-layer baryonic matter from the vacuum occurs at exactly the same point as the onset of 2-layer baryonic matter, but 1-layer baryonic matter is energetically disfavored from this point on, although close to the onset it is barely distinguishable  from 2-layer baryonic matter. Therefore, and since we ignore the multi-layer solutions for the reasons explained above, the 2-layer configuration is the only way we add baryons to the system.

Next we need to discuss the phases where the flavor branes reach the horizon at $u=u_T$. In this case, instead of stationarity with respect to $u_c$, we must require stationarity with respect to $x_4(u_T)$. Instead of Eqs.\ (\ref{min1}) we derive
\begin{subequations}\allowdisplaybreaks\label{min1H}
\bea
\frac{\ell}{2} &=&   \int_{u_T}^\infty du\,x_4'+x_4(u_T) \, , \label{x4uT} \\[2ex]
\int_{u_T}^{u_b}du\,\hat{a}_0'&=&\frac{u_b}{3}\sqrt{f_T(u_b)}  \, ,\label{MixCond1} \\[2ex]
\int_{u_b}^\infty du\, \hat{a}_0'&=&\mu- \frac{u_b}{3}\sqrt{f_T(u_b)} \, , \label{MuMix1}\\[2ex]
k&=& -A\phi_T(u_T) \, , \label{kA}\\[2ex] 
n &=& \frac{6\Delta_T(u_b)}{u_b^{3/2}} \sqrt{\frac{
u_b^8 f_T(u_b) - [A \phi_T(u_b) + k]^2 }{
f_T(u_b) [9-\Delta_T^2(u_b)] [9 (2-x_b)^2-x_b^2\Delta_T^2(u_b)]
}} \, , \label{nMix1}
\eea  
\end{subequations}
plus the equation for $A$, which now takes the form
\be \label{eqAH}
A = \frac{2\al}{\lambda_0^2}
    \exp \left[ 2\lambda_0\left( \frac{\ell}{2} \phi_T(u_T) +  \int_{u_T}^\infty du\,  \left[\phi_T(u) - \phi_T(u_T) \right] x_4'(u)\right) \right] \, .
\ee  
This generalizes the phase introduced in Ref.\ \cite{Kovensky:2019bih} by allowing pointlike instantons to sit at $u_b>u_T$, see right panel of Fig.\ \ref{fig:quarkyonic}. As discussed in Ref.\ \cite{Kovensky:2019bih}, in the case of a nonzero current quark mass the embedding of the flavor branes is non-straight, it approaches the horizon tangentially, and the geometry only exists at nonzero temperatures. In the chiral limit, the branes become straight, $k=0$, and the geometry exists for all temperatures. As for the above case with branes joining at $u_c$, the source terms do not contribute explicitly to the free energy at the stationary point, which, analogously to Eq.\ (\ref{Omstat1}), becomes,
\be\label{Omstat2}
\Omega = \int_{u_T}^\infty du\, u^{5/2}\zeta(u) -\frac{A}{2\lambda_0} \, .
\ee
In this geometry, the quark density is always nonzero 
(unless $\mu=0$) and thus we have the following two phases.

\begin{enumerate}

\item[$(e)$] {\it High-temperature quark phase (HTQ): $n_q>0,n_b=0$.}

In this case, the baryon density is purely generated by quarks,
and the relevant equations that need to be solved are 
(\ref{x4uT}), (\ref{minnq}), (\ref{kA}), and (\ref{eqAH}). 
(Again, as discussed for the LTQ phase $(c)$, we need to go back to the stationarity condition with respect to $n_q$ (\ref{minnq}), since Eq.\ (\ref{MixCond1}) only holds in the presence of baryons.) We can add baryons to this phase continuously, but as for the low-temperature counterpart there is in general no simple analytical expression for the critical chemical potential of this transition. Such an expression can be derived, however, in the chiral limit, see Sec.\ \ref{sec:chiral}.

 \item[$(f)$] {\it High-temperature quarkyonic phase (HTQy): $n_q,n_b>0$.} 
 
 This is the high-temperature counterpart of the LTQy phase $(d)$, see right panel of Fig.\ \ref{fig:quarkyonic}. In this case, we need to solve all equations (\ref{min1H}) and (\ref{eqAH}) simultaneously. As discussed in Ref.\ \cite{Kovensky:2019bih}, the existence of two distinct geometries for nonzero quark density may be due to a simplified description using strings rather than a more complicated structure of the flavor branes. This is supported by the observation that, in the massive case, where a transition between the LTQ and HTQ phases takes place, this transition is "nearly" a crossover, with a very small discontinuity in the speed of sound \cite{Kovensky:2019bih}. This is also true for the present generalization to the quarkyonic phases, such that for the physical interpretation it is, for most purposes, useful to think of the LTQy and HTQy geometries as describing the same physical phase. 
  
 \end{enumerate}
 
Our holographic quarkyonic phases, LTQy $(d)$ and HTQy $(e)$, are novel configurations, never discussed before in the Witten-Sakai-Sugimoto model. As we shall see, they cover a significant part of the phase diagram. Therefore, and independent of their interpretation, one of our main results is to point out that they have to be taken into account in the evaluation of the model at nonzero densities and chemical potentials. Before 
discussing the phase structure in detail, 
let us comment on our interpretation of these phases, i.e., on why we term them "quarkyonic". As explained in the introduction, the quarkyonic phase of large-$N_c$ QCD can be characterized by various properties. Our LTQy  and HTQy phases are a simple version of quarkyonic matter, and we do not claim to have constructed a holographic equivalent that shares all characteristics with the quarkyonic phase of large-$N_c$ QCD or possibly of real-world QCD. Let us first recall the physical picture of the quarkyonic phase in large-$N_c$ QCD \cite{McLerran:2007qj,McLerran:2018hbz,Philipsen:2019qqm}: in momentum space (and at zero temperature), this phase can be thought of as a Fermi sea of quarks on top of which there is a layer of baryons. 
If the Fermi momentum is large compared to the width of the baryon layer, the thermodynamics are dominated by quarks, although the  excitations, relevant for instance for transport properties, are  baryonic. In this picture, the quarkyonic phase is not unlike a fermionic system with Cooper pairing in the vicinity of the Fermi surface. Increasing the baryon chemical potential, one expects to go from a baryonic phase (purely baryonic Fermi sea) through a quarkyonic phase (in which a quark Fermi sea starts to grow) to a quark phase (where the layer of baryons has disappeared). In particular, there are two contributions to the baryon density, one from actual baryons, and one from quarks. This picture has recently been translated into a simple phenomenological model for quarkyonic matter \cite{McLerran:2018hbz}. At strong coupling, a sharp Fermi surface is not expected (although some elements of Fermi liquid behavior have been observed in holographic calculations \cite{Karch:2008fa,Kulaxizi:2008jx,DiNunno:2014bxa}).
Therefore, our holographic quarkyonic phase is perhaps best understood as a strong-coupling version of the model introduced in Ref.\ \cite{McLerran:2018hbz}, not relying on the presence of any Fermi surface. We also have contributions from quarks and baryons to the density, and the Fermi sea construction is replaced by the geometries shown in Fig.\ \ref{fig:quarkyonic}. Since the holographic coordinate $u$ is interpreted as an energy scale we may assign different energy regimes to quarks and baryons, reminiscent of the momentum-space picture. We shall elaborate on our holographic picture of quarkyonic matter in Sec.\ \ref{sec:embed}, where we show how the geometry develops as temperature and chemical potentials are varied.  

To avoid any confusion, we emphasize that the holographic quarkyonic phase (just like the large-$N_c$ QCD version) is not a "mixed" phase of quarks and baryons. In a mixed phase, according to the usual terminology, two phases are spatially separated or possibly form a homogeneous mixture if they  interpenetrate each other, like a mixture of two gases. In the weak-coupling picture, this would correspond to two {\it separate} Fermi surfaces for quarks and baryons. In our model we could construct such a mixed phase by assigning different volume fractions (to be determined dynamically) to the pure quark phase and the pure baryonic phase. It is conceivable that such a phase is preferred in a certain region of the phase diagram, in particular close to a first-order phase transition between the two pure phases, as discussed routinely in the context of dense matter in neutron stars \cite{Glendenning:1992vb,Heiselberg:1992dx,Schmitt:2020tac}. In this work we ignore such a mixed phase, but emphasize that the quarkyonic phase is qualitatively different. 

Since in the usual picture of the quarkyonic phase excitations at the Fermi surface are color singlets, the quarkyonic phase is usually argued to be confined. It is tempting to make the connection to our holographic version by viewing our quarkyonic (and baryonic) phases as being confined. However, as explained in Sec.\ \ref{sec:setup}, we work in the "deconfined geometry", and thus have to be careful with this interpretation. One might argue that the deconfined geometry only refers to deconfined gluons, which are described by the background geometry and which we have decoupled from our description, and that quarks can nevertheless be "confined" in baryons. We do not attempt to make these arguments regarding confinement more precise. What we {\it can} do in a precise sense is to observe spontaneous chiral symmetry breaking (or absence thereof)  within the quarkyonic phase, at least in the  limit of a vanishing current quark mass, where chiral symmetry is exact. We have constructed two realizations of quarkyonic matter which, in the massless limit, correspond to chirally restored quarkyonic matter (HTQy, flavor branes disconnected) and chirally broken quarkyonic matter (LTQy, flavor branes connected). The system will choose the preferred configuration dynamically, and we will discuss the result in Sec.\ \ref{sec:thermo}.

Finally, we should mention that in large-$N_c$ QCD quarkyonic matter has first been identified by power counting in $N_c$. It was realized that there is a phase at large densities and sufficiently low temperature whose pressure scales as $N_c$ -- suggesting quark degrees of freedom to be dominant --  although the phase is confined. This is in contrast to the pressure of the mesonic and the purely baryonic phases, which are expected to scale as $N_c^0$, and the deconfined quark-gluon plasma, which scales as $N_c^2$ because it is dominated by gluons. Our model in its present form does not allow us to reproduce these scalings. Firstly, since we work in the deconfined geometry, the gluonic background always gives an $N_c^2$ contribution, such that strictly speaking the pressure of all our phases is dominated by gluons. Of course, in the spirit of the decompactified limit, we ignore this contribution and may still ask whether our phases scale differently with $N_c$. However, taking for instance Eq.\ (\ref{Omega}) together with the definition of ${\cal N}$ (\ref{N}) we see that the pressure (which is the negative of the free energy density) always scales like $N_fN_c$. The same observation has been made in a D3/D7 model, see Ref.\ \cite{Mateos:2007vn} for a discussion of the scaling properties in a probe brane setup at strong coupling. Thus we cannot use naive power counting in $N_c$ in the present approach to distinguish between our phases.

We now turn to the evaluation of the stationarity equations and the corresponding free energies. In general, this has to be done  numerically. In the chiral limit some analytical results can be derived, as we now explain. 

\subsection{Chiral limit}
\label{sec:chiral}

In the chiral limit, it turns out that the LTQy phase is metastable at best, just like the LTQ phase in the absence of baryon sources  \cite{Kovensky:2019bih}. We can therefore focus on the HTQy phase and thus on Eqs.\ (\ref{min1H}), where we can set $A=0$. It follows immediately from Eq.\ (\ref{kA}) that $k=0$ and thus with Eq.\ (\ref{solx4}) that the flavor branes are straight, $x_4'(u) =0$, for all $u$. In the remaining equations all integrals can be performed, and Eqs.\ (\ref{MixCond1}), (\ref{MuMix1}), and (\ref{nMix1}) read 
\begin{subequations} \label{chiral}
\begin{eqnarray}
    \frac{\sqrt{f_T(u_b)}}{3} &=& 
    {}_2F_1 \left[
    \frac{1}{5},\frac{1}{2},\frac{6}{5},-\frac{u_b^5}{n^2(1-x_b)^2}\right]
    - \frac{u_T}{u_b}
    {}_2F_1 \left[
    \frac{1}{5},\frac{1}{2},\frac{6}{5},-\frac{u_T^5}{n^2(1-x_b)^2}\right] \, ,\label{chiral2}\\[2ex]
    \mu &=& 
    u_b \left\{
    \frac{\sqrt{f_T(u_b)}}{3} - {}_2F_1 \left[
    \frac{1}{5},\frac{1}{2},\frac{6}{5},-\frac{u_b^5}{n^2}\right]\right\}+ n^{2/5} \frac{\Gamma[3/10]\Gamma[6/5]}{\sqrt{\pi}}
    \, , \label{chiral1}\\[2ex]
    n &=& \frac{6 \, u_b^{5/2}  \Delta_T(u_b)} {\sqrt{ [9-\Delta_T^2(u_b)] [9 (2-x_b)^2-x_b^2\Delta_T^2(u_b)]
}} \, .  \label{chiral3}
\label{ntotx0}
\end{eqnarray}
\end{subequations}
Given $\mu$ and $t$, these are three coupled equations for $n$, $u_b$, and $x_b$. The dimensionless free energy density  
can be computed from Eq.\ (\ref{Omstat2}), which, 
after subtracting the vacuum contribution and using Eqs.\ (\ref{chiral2}) and (\ref{chiral1}), can be written as
\bea \label{Om1}
     \Omega^{\rm HTQy}_{m_q=0}(\mu,t)  &=& -\frac{2}{7}
    \left\{n \mu + u_T \sqrt{n^2(1-x_b)^2 + u_T^5} \right. \non[2ex]
    && \left. - u_b \left[
    \sqrt{n^2(1-x_b)^2+u_b^5}-
    \sqrt{n^2+u_b^5} + 
    \frac{nx_b}{3}\sqrt{f_T(u_b)} 
    \right]\right\} \, .
\eea
For comparison, in the pure quark phase in the chiral limit, the only nontrivial equation is
\be
\mu= \int_{u_T}^\infty du\, \hat{a}_0' =  -u_T{}_2F_1\left[\frac{1}{5},\frac{1}{2},\frac{6}{5},-\frac{u_T^5}{n^2}\right]+\frac{n^{2/5}\Gamma[3/10]\Gamma[6/5]}{\sqrt{\pi}} \, ,
\ee
and the free energy density can be written as
\be \label{OmHTQ}
\Omega^{\rm HTQ}_{m_q=0}(\mu,t) = -\frac{2}{7}\left(n\mu+u_T\sqrt{n^2+u_T^5}\right) \, .
\ee
By taking the limit $x_b \to 0$ in Eqs.\ (\ref{chiral}) we obtain the onset of quarkyonic matter from pure quark matter. We find
\begin{equation} \label{uT3}
 u_b = c_1 u_T \, ,
 \qquad 
 \mu = c_2 u_T \, ,
 \qquad 
 n = c_3 u_T^{5/2} \, ,
\end{equation}
where $c_1 \simeq 1.92863$ satisfies 
\begin{equation}
    \frac{\sqrt{c^3-1}}{3c^{1/2}} = c \, 
    {}_2F_1 \left[
    \frac{1}{5},\frac{1}{2},\frac{6}{5},-g(c)\right]
    - 
    {}_2F_1 \left[
    \frac{1}{5},\frac{1}{2},\frac{6}{5},-\frac{g(c)}{c^5}\right] \, , 
\end{equation}
with $g(c) \equiv
(32c^6 - 40c^3 - 1)/(1+2c^3)^{2}$, and 
\begin{subequations}
\bea
    c_2 &=& - \,
    {}_2F_1 \left[
    \frac{1}{5},\frac{1}{2},\frac{6}{5},-\frac{g(c_1)}{c_1^5}\right] + \frac{c_1}{g^{1/5}(c_1)}
    \frac{\Gamma[3/10]\Gamma[6/5]}{\sqrt{\pi}}\simeq  1.12171 \, , \\[2ex] 
    c_3 &=& \frac{c_1^{5/2}}{g^{1/2}(c_1)} \simeq 2.15067\, .
\eea
\end{subequations}
With the help of Eq.\ (\ref{tuT}), the second relation in Eq.\ (\ref{uT3}) gives the following curve in the $t$-$\mu$ plane,
\begin{equation}
t = \frac{3}{4 \pi} \sqrt{\frac{\mu}{c_2}} \, .    
\label{tmusqrt}
\end{equation}
In the full numerical evaluation (including all other candidate phases) we shall see that this onset is realized in the phase diagram for sufficiently large temperatures. The relation (\ref{tmusqrt}) shows that the critical temperature of the quark-quarkyonic transition {\it increases} with $\mu$. This is perhaps surprising, given that this transition is constant in $\mu$ for $N_c=\infty$ QCD, i.e., given by a horizontal line in the $t$-$\mu$ plane, which is usually believed to bend downwards, not upwards, as $N_c$ is decreased \cite{McLerran:2007qj}. 

At zero temperature, we can further simplify the result. In this case, $u_T=0$, $f_T(u)=1$, and $\Delta_T(u_b)=1$. Therefore, inserting Eq.\ (\ref{chiral3}) into Eq.\ (\ref{chiral2}) yields an equation for $x_b$ only, 
\begin{equation}
    {}_2 F_1 \left[\frac{1}{5},\frac{1}{2},
    \frac{6}{5},-\frac{2[9 (2-x_b)^2-x_b^2]}{9(1-x_b)^2}\right] = \frac{1}{3}\, , 
    \label{xmq0}
\end{equation}
with the (only relevant) numerical solution 
\be
x_b \simeq 0.966471 \, . 
\ee
Strikingly, the baryon fraction is very close to 1, i.e., in terms of density contributions our holographic quarkyonic phase is almost purely baryonic at zero temperature, at least in the chiral limit. The suppression of the quark contribution is in accordance with the model of Ref.\ \cite{McLerran:2018hbz}, where the quark density in the quarkyonic phase is suppressed parametrically by $1/N_c^3$ compared to the density from baryons\footnote{ Amusingly, if this suppression is taken literally, i.e., without any additional numerical factors, the result for $N_c=3$ is very close to the value we found, $1-1/3^3\simeq 0.962963$.}. It is also interesting, and perhaps surprising, that the baryon fraction is constant, i.e., it does not depend on $\mu$. In particular, we can already conclude that in the chiral limit and at zero temperature there cannot be any continuous transition from the quarkyonic phase to either the pure quark phase (where $x_b=0$) or the pure baryonic phase (where $x_b=1$).

The remaining equations (\ref{chiral1}) and (\ref{chiral3}) give 
\be
n=X_2 \mu^{5/2} \, , \qquad u_b = X_1^{1/5} X_2^{2/5} \mu \, ,
\ee
with 
\begin{subequations}
\bea
X_1&\equiv& \frac{2[9(2-x_b)^2-x_b^2]}{9} \, , \\[2ex]
X_2 &\equiv& \left\{\frac{\Gamma[3/10]\Gamma[6/5]}{\sqrt{\pi}}+X_1^{1/5}\left(\frac{1}{3}-{}_2F_1\left[\frac{1}{5},\frac{1}{2},\frac{6}{5},-X_1\right]\right)\right\}^{-5/2} \, .
\eea
\end{subequations}
Inserting all this into the free energy density (\ref{Om1}) yields
\be \label{OmMix00}
\Omega^{\rm HTQy}_{m_q=0}(\mu,t=0) = -\frac{2}{7}X_2\mu^{7/2}\simeq - 0.371413 \mu^{7/2} \, . 
\ee
This is to be compared to the free energy density (\ref{OmHTQ}) of the pure quark phase, which becomes
\be \label{OmHTQ00}
\Omega^{\rm HTQ}_{m_q=0}(\mu,t=0) = -\frac{2}{7}\left(\frac{\Gamma[3/10]\Gamma[6/5]}{\sqrt{\pi}}\right)^{-5/2} \mu^{7/2}\simeq - 0.0955687 \mu^{7/2} \, .
\ee
We observe, first of all, that both quark and quarkyonic phases exhibit the same $\mu^{7/2}$ behavior. This is remarkable because we have just seen that the density of the quarkyonic phase is utterly dominated by baryons, and yet it seems to behave in some sense like quark matter. For instance, as a consequence, the zero-temperature speed of sound of quarkyonic matter and quark matter in the chiral limit are exactly the same, see discussion at the end of Sec.\ \ref{sec:continuity}. Furthermore, Eqs.\ (\ref{OmMix00}) and (\ref{OmHTQ00}) show that the quarkyonic phase is preferred over the pure quark phase for all $\mu$. (At small $\mu$, as expected and as we shall see, the mesonic and purely baryonic phases become favored over both phases.) Therefore, 
the quarkyonic configuration must be included in the phase structure of the model. And we conclude that, in contrast to QCD, pure quark matter does not become the ground state at asymptotically large $\mu$. This is not too surprising since our present model is not asymptotically free, such that comparisons with QCD become meaningless at ultra-high chemical potentials and/or temperatures. Moreover, at large chemical potentials, our probe brane approximation cannot be expected to be reliable, and backreactions should be taken into account. It is then conceivable that pure quark matter does become favored at large chemical potentials, as seen for instance in Ref.\ \cite{Ishii:2019gta}.

\section{Numerical results and discussion}
\label{sec:thermo}

In this section we present the numerical results, obtained by solving the coupled system of algebraic equations (containing numerical integrals) presented in the previous section, i.e., Eqs.\ (\ref{min1}) and (\ref{eqA}) for the phases $(a)$, $(b)$, $(c)$, $(d)$, and Eqs.\ (\ref{min1H}) and (\ref{eqAH}) for the phases $(e)$, $(f)$. The solutions are then inserted back into the free energy density (\ref{Omstat1}) and (\ref{Omstat2}), respectively, to determine the favored phase for all temperatures and chemical potentials.

We can eliminate  the asymptotic separation of the flavor branes $\ell$ from our equations by rescaling all dynamical and thermodynamic variables by appropriate powers of $\ell$. In all figures the results will therefore be presented in terms of these rescaled (and dimensionless) quantities. The relevant rescaled thermodynamic quantities are $\tilde{t} = t\ell$, $\tilde{\mu} = \mu\ell^2$, $\tilde{n} = n\ell^5$, $\tilde{\Omega} = \Omega\ell^7$, while the baryon fraction $x_b$ obviously does not get rescaled. We shall also plot the embedding of the flavor branes and the abelian gauge field, for which we need $\tilde{u} = u\ell^2$, $\tilde{x}_4=x_4/\ell$, and $\tilde{a}_0=\hat{a}_0\ell^2$. This leaves us with only two parameters which we have to specify for the calculation, the rescaled 't Hooft coupling $\tilde{\lambda}=\lambda/\ell$ and the rescaled mass parameter $\tilde{\alpha} = \alpha \ell^4$. (If $\tilde{\alpha}=0$, then  $\tilde{\lambda}$ does not appear explicitly in the calculation.) We shall work with the fixed value $\tilde{\lambda} = 15$. This value is motivated by a fit to properties of nuclear matter at saturation, obtained from a more refined, non-pointlike, instanton approximation \cite{BitaghsirFadafan:2018uzs}. It is also comparable to the value obtained by fitting the model parameters to the pion decay constant and the rho meson mass in the original version of the model \cite{Sakai:2005yt} if we assume $\ell\sim 1$, corresponding to asymptotically separating the flavor branes by about a third of the antipodal separation. For a systematic study of how the 't Hooft coupling changes the phase structure in the current setup -- but without baryons -- see Ref.\ \cite{Kovensky:2019bih}. We will consider different values of the mass parameter $\tilde{\alpha}$ in order to study the effect of a nonzero pion mass. Here we have to keep in mind that going to large values of the pion mass is an extrapolation because we have neglected higher-order terms in the current quark mass from the beginning. 

We start with a discussion of the $\tilde{t}$-$\tilde{\mu}$ phase diagram and then elaborate on certain aspects of this diagram related to quarkyonic matter. 

\subsection{Quarkyonic matter in the phase diagram}
\label{sec:phasediagram}

\begin{figure}[t]
    \centering
        \includegraphics[width=0.49\textwidth]{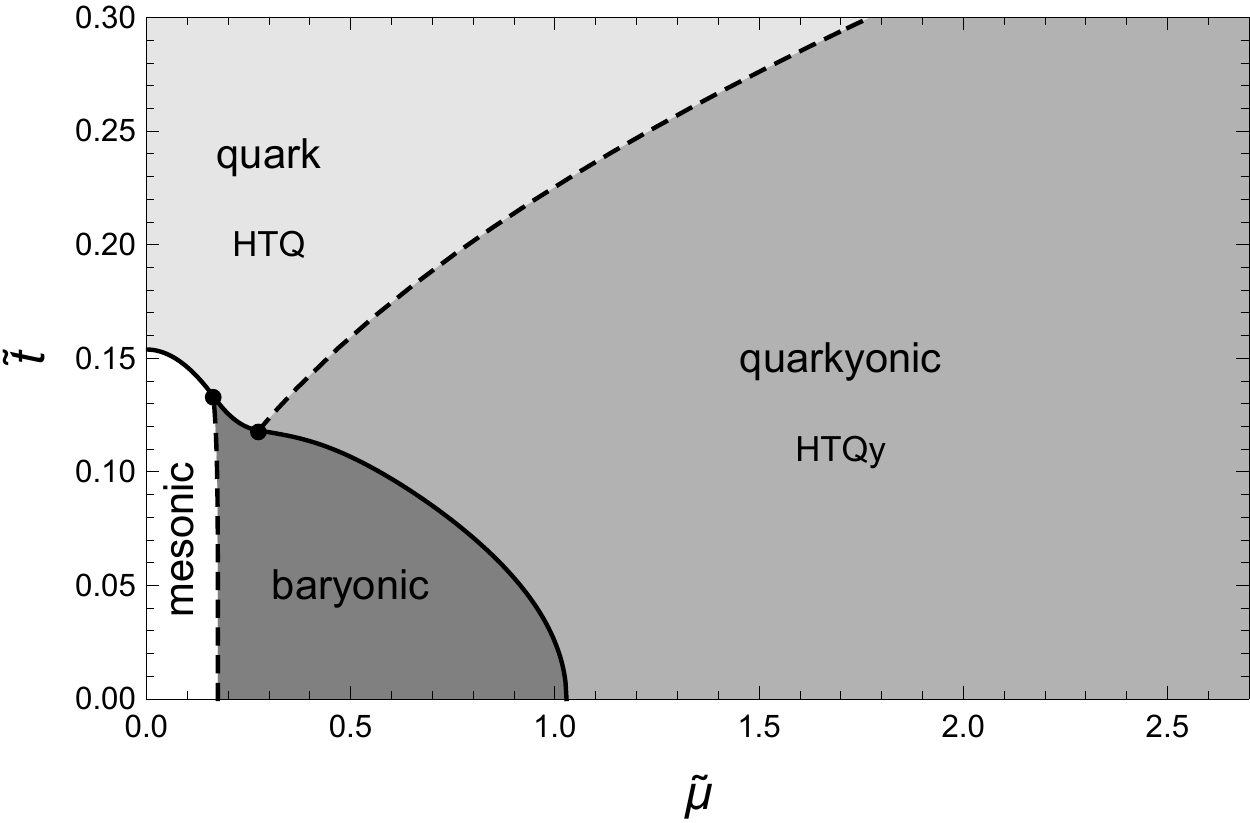}
        \includegraphics[width=0.49\textwidth]{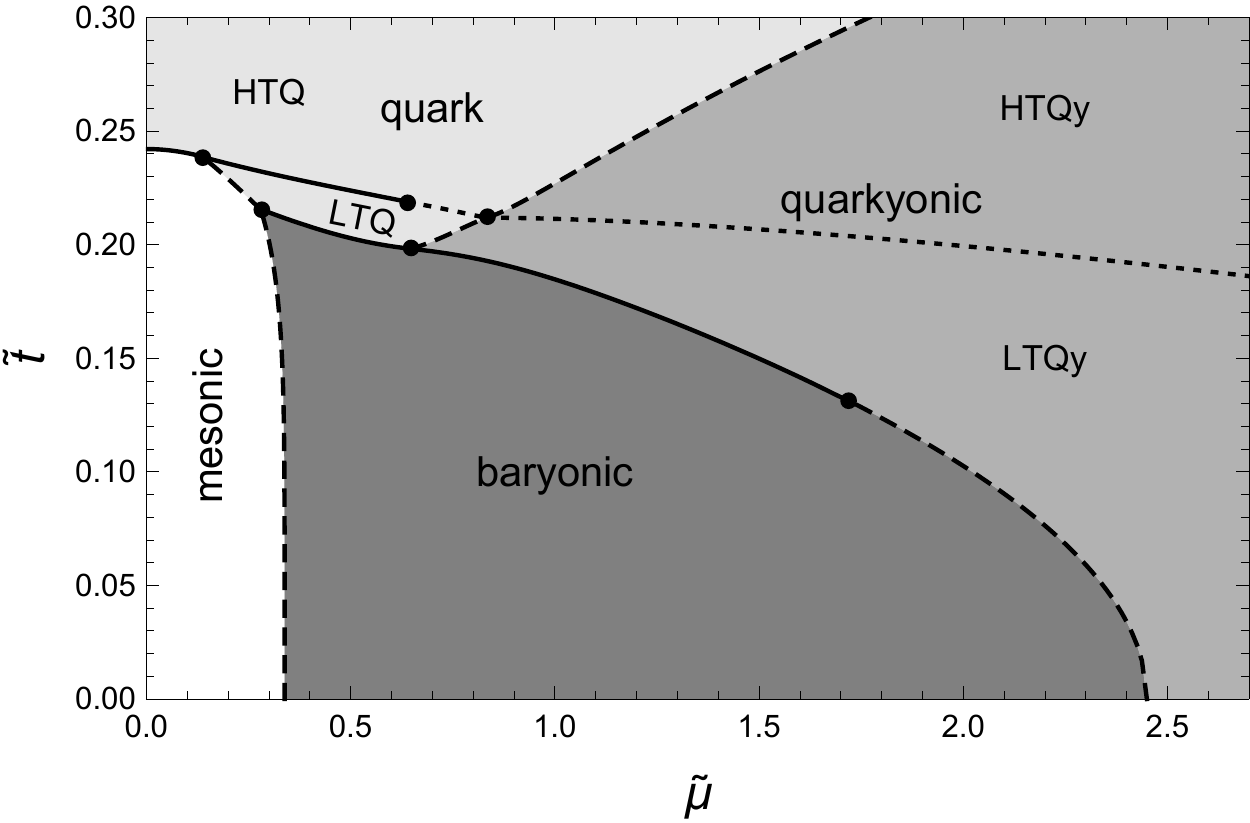}
        \caption{{\small Phase diagrams in the plane of dimensionless temperature and quark chemical potential for 't Hooft coupling $\tilde{\lambda}=15$ and quark mass parameters $\tilde{\alpha}=0$ (chiral limit, left panel) and $\tilde{\alpha} = 0.2$ (large pion mass, right panel). Solid lines are discontinuous phase transitions while transitions across dashed lines are continuous. In the right panel, the continuous transition between the two different versions of pure quark matter (HTQ and LTQ) and the two versions of quarkyonic matter (HTQy and LTQy) is marked by a dotted line. }}
         \label{fig:phasediagrams}
\end{figure}

In Fig.\ \ref{fig:phasediagrams} we present the phase diagram in the chiral limit (left panel, $\tilde{\alpha}=0$) and for a nonzero quark mass parameter (right panel, $\tilde{\alpha}=0.2$). Our observations are as follows. 

\begin{itemize}

\item At low temperature, by increasing the chemical potential we move from the vacuum through baryonic matter to the quarkyonic phase. As already observed from the analytical results in Sec.\ \ref{sec:chiral}, there is no high-density transition to pure quark matter at low temperatures. Baryons in the quarkyonic phase do, however, melt as the temperature is increased, and pure quark matter is reached through a continuous transition. This is discussed in more detail in Sec.\ \ref{sec:frac}. 

\item In the quarkyonic phase, chiral symmetry is restored. 
In the massless scenario of the left panel,  the flavor branes  are straight and disconnected in the  quarkyonic phase (HTQy), and thus chiral symmetry is intact. A chirally broken branch (LTQy) does exist even in the chiral limit as a stationary point of the free energy but is never preferred. As a consequence, allowing for quarkyonic matter has introduced a zero-temperature chiral phase transition into the phase diagram, which otherwise would be absent in the approximation of pointlike baryons \cite{Bergman:2007wp}. For a small explicit chiral symmetry breaking, the quarkyonic phase is "almost" chirally symmetric, because the stable branches of both HTQy and LTQy configurations have nearly straight branes.

\item For large values of the pion mass, the low-temperature transition from baryonic to quarkyonic matter becomes continuous. By taking a suitable path in the phase diagram we can even connect all possible phases continuously. Here, by continuously we mean via phase transitions where the first derivatives of the thermodynamic potential (i.e., density and entropy) are continuous, although higher derivatives may be discontinuous. We will comment on this continuity in more detail in Sec.\ \ref{sec:continuity}.

\item At nonzero pion mass, the quark and quarkyonic phases each have a low-temperature and a high-temperature version. We have denoted the transitions between them with dotted lines, suggesting that we do not interpret them as physically distinct phases (which is also suggested by the shading of the phases). It is conceivable that this transition becomes a true crossover in a more refined approximation, as already discussed in Ref.\ \cite{Kovensky:2019bih}.

\item While in QCD at infinite $N_c$ mesonic, quarkyonic, and quark phases are expected to meet in a triple point \cite{McLerran:2007qj}, this structure is more complicated in our model already in the chiral limit. The reason is the presence of the pure baryonic phase, which we know exists in  real-world QCD, and we observe two triple points. At large pion masses, there are still two triple points, but an additional discontinuous transition within the quark phase has occurred, which was already observed in Ref.\ \cite{Kovensky:2019bih}. 

\item The baryon onset (i.e., the  transition from the mesonic to the baryonic phase) is continuous, and thus there is no first-order liquid-gas transition as in real-world QCD. This is a known unphysical feature of the pointlike approximation, already observed for 1-layer baryonic matter in the chiral limit \cite{Bergman:2007wp}. This problem can be cured by allowing for instantons with nonzero widths \cite{Preis:2016fsp,BitaghsirFadafan:2018uzs} or by using a different approximation for the non-abelian gauge fields that is not directly based on instantons \cite{Rozali:2007rx,Li:2015uea}.

\end{itemize}

Let us support these qualitative observations with some quantitative estimates. To this end, we need to connect our dimensionless variables to physical units. 
We use Eqs.\ (\ref{defalpha}) and   (\ref{fpi}) to write pion mass and pion decay constant as 
\be
m_\pi^2 \simeq 109.2\, \frac{\tilde{\alpha}}{\tilde{\lambda}L^2}\,  e^{\frac{\tilde{\lambda}}{4}\tan\frac{\pi}{16}} \, , \qquad f_\pi^2 \simeq 3.093\times 10^{-4} \frac{N_c\tilde{\lambda}}{L^2} \, . \label{mpi}
\ee 
With $N_c=3$, $\tilde{\lambda}=15$,  and $f_\pi\simeq 93\, {\rm MeV}$ this yields $L\simeq (790\, {\rm MeV})^{-1} \simeq 0.25 \, {\rm fm}$. Consequently, we find that $\tilde{\alpha} \simeq 2.05\times 10^{-3}$ corresponds to the physical pion mass $m_\pi\simeq 140\, {\rm MeV}$, while $\tilde{\alpha}=0.2$, chosen for the right panel of Fig.\ \ref{fig:phasediagrams}, corresponds to an unphysically large pion mass $m_\pi \simeq 1.38 \, {\rm GeV}$. (This somewhat improves the similar estimate of Ref.\  \cite{Kovensky:2019bih}, where $f_\pi$ was not fitted separately and an arbitrary value for $L$ was assumed, leading to a slightly smaller $\tilde{\alpha}$ at the physical point.) There are two main reasons why we vary the pion mass far beyond its physical value. Firstly, in doing so we explore the parameter space to look for qualitatively different scenarios. Although they might be associated to unphysically heavy quarks in the given model, it cannot be excluded that they are realized in nature. Secondly, studying "heavy holographic QCD" with baryonic and quarkyonic matter is interesting in view of comparisons to lattice calculations at nonzero baryon densities, which can be performed with the help of an expansion about large quark masses \cite{Philipsen:2019qqm}. 

We may also use the numerical value of $L$ to give an estimate of the important transition points in the phase diagrams. To this end, we employ the relations of Table 1 in Ref.\ \cite{Kovensky:2019bih} to compute the temperature, $\tilde{t}/L$, and the quark chemical potential, $\tilde{\lambda}\tilde{\mu}/(4\pi L)$,
from the dimensionless, rescaled variables. For instance, from the left panel of Fig.\ \ref{fig:phasediagrams} we find that in the chiral limit the zero-density chiral transition is at $T_c\simeq 120\, {\rm MeV}$, and the zero-temperature onset of baryonic matter is at $\mu_0\simeq 165\, {\rm MeV}$. These numbers change only very slightly if evaluated at the physical point $\tilde{\alpha} \simeq 2.05\times 10^{-3}$, see for instance  Fig.\ 6 of Ref.\ \cite{Kovensky:2019bih} for the effect of $\tilde{\alpha}=10^{-3}$ on the chiral phase transition in the absence of baryons.  In real-world QCD, the baryon onset is at $\mu_0\simeq 308\, {\rm MeV}$, slightly less than the baryon mass divided by $N_c$ due to the binding energy of nuclear matter. Of course, within our pointlike approximation we do not expect to reproduce realistic nuclear matter (and we have already seen that this approximation predicts an unrealistic second-order baryon onset). Moreover, we have set $\tilde{\lambda}=15$ by hand and could in principle include $\mu_0$ into our fitting procedure to adjust $\tilde{\lambda}$. Nevertheless, since the value we have used is motivated by the more realistic approximation of Ref.\ \cite{BitaghsirFadafan:2018uzs}, where the fit {\it did} include nuclear matter properties, but $f_\pi$ was {\it not} used, our estimates suggest that there is some tension in the model between reproducing vacuum properties such as the pion decay constant and properties of nuclear matter. 

With these caveats in mind, the most interesting estimate for our present purposes concerns the transition to quarkyonic matter. We find that 
at zero temperature it occurs at a quark chemical potential $\mu_c\simeq 970\, {\rm MeV}$ in the chiral limit and at $\mu_c\simeq 2.3\, {\rm GeV}$ for $m_\pi\simeq 1.38 \, {\rm GeV}$ (the corresponding baryon chemical potentials are $N_c\mu_c$). 
Therefore, even in the chiral limit the  chiral phase transition, i.e., the transition from baryonic to quarkyonic matter,  occurs at a very large chemical potential compared to the interior of neutron stars, where we expect quark chemical potentials of up to about $500\, {\rm MeV}$. At nonzero temperatures, however, we see that the quarkyonic transition occurs at smaller chemical potentials. For instance, we read off that at a quark chemical potential of $500\, {\rm MeV}$ the transition occurs at about $T_c\simeq 80 \, {\rm MeV}$. This temperature is possibly reached in neutron star mergers \cite{Most:2019onn}. Therefore, in a merger the appearance of quarkyonic matter seems possible, but in isolated neutron stars, where the temperature is essentially zero on the scale shown here, we would expect the core to be purely baryonic. It is likely, however, that this conclusion is altered by a more realistic approximation of baryons. The reason is that if nuclear matter is described by instantons with nonzero width
and quarkyonic matter is ignored, there is a zero-temperature chiral transition to quark matter \cite{Li:2015uea,BitaghsirFadafan:2018uzs}, which is completely absent in the pointlike approximation \cite{Bergman:2007wp}. This suggests that more realistic baryons are less favored compared to chirally symmetric matter than pointlike baryons. Hence it would be interesting to see whether in such a more realistic treatment the baryonic-quarkyonic transition moves to lower chemical potentials and whether a subsequent quarkyonic-quark transition is introduced at low temperatures.

\subsection{Baryon fraction in the quarkyonic phase}
\label{sec:frac}

\begin{figure}[t]
    \centering
        
       \includegraphics[width=0.49\textwidth]{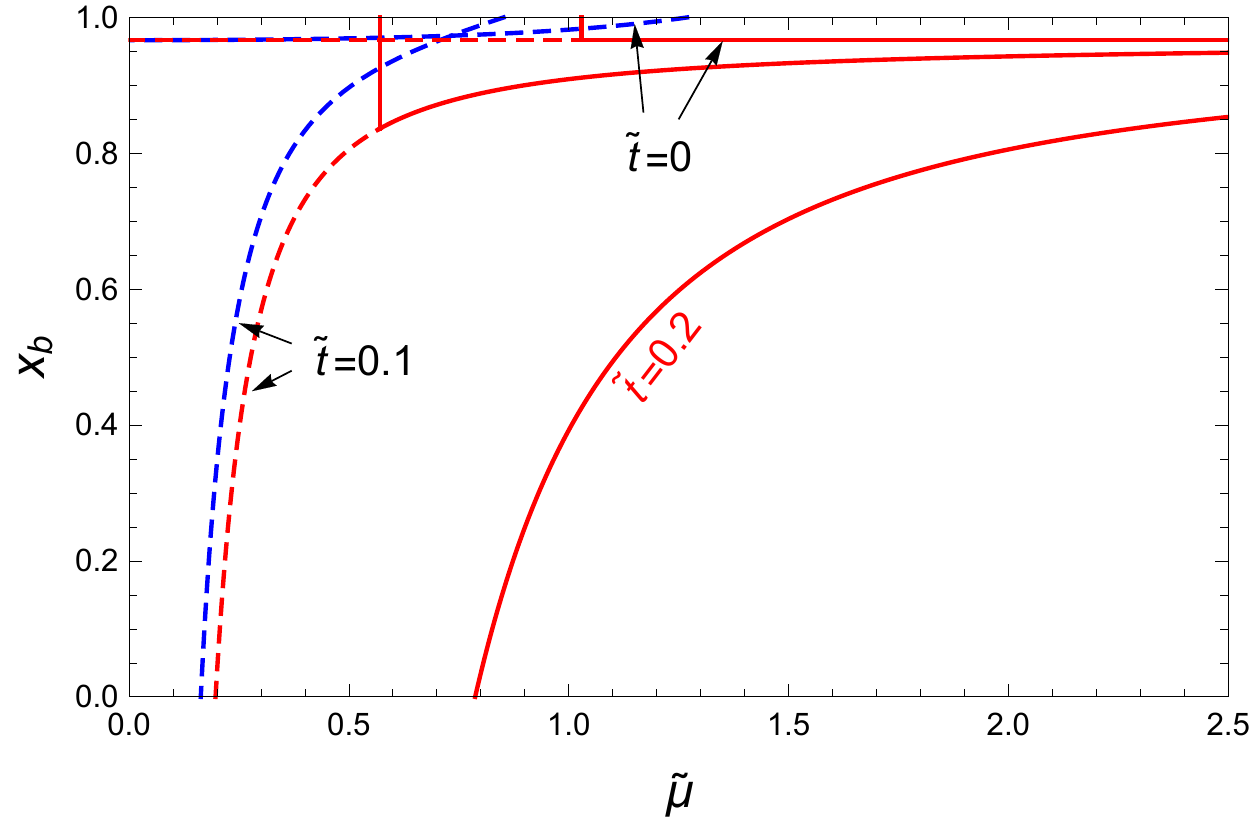} \includegraphics[width=0.49\textwidth]{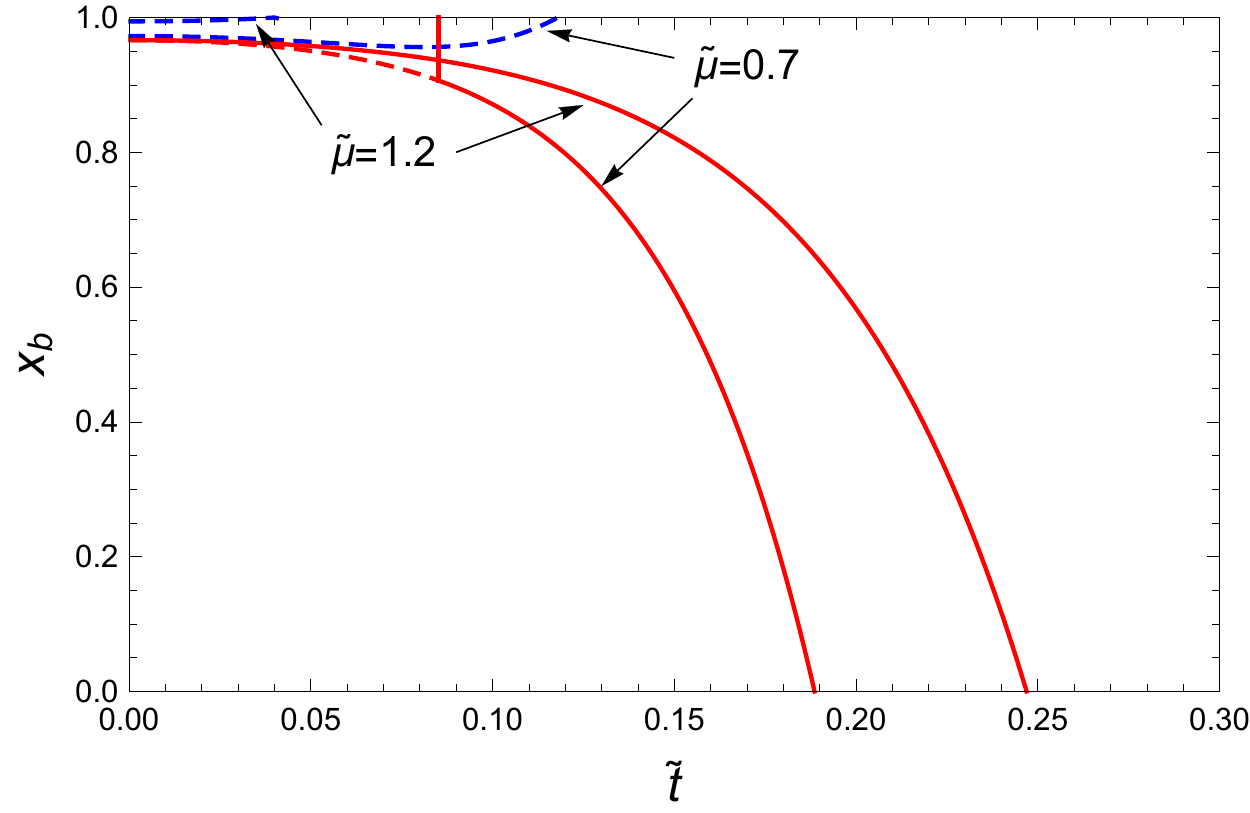}

         \caption{{\small Density fraction from baryons $x_b$ in  quarkyonic matter in the chiral limit, $\tilde{\alpha}=0$, corresponding to the phase diagram in the left panel of Fig.\ \ref{fig:phasediagrams}, for three different temperatures as a function of chemical potential  (left) and for two different chemical potentials as a function of temperature (right).
         Solid lines correspond to the stable phases, including discontinuous transitions from the purely baryonic phase ($x_b=1$). All stable segments are in the HTQy configuration. Dashed curves correspond to unstable quarkyonic branches in the HTQy (red) and LTQy (blue) configurations.}}
         \label{fig:xbchiral}
\end{figure}
\begin{figure}[ht]
    \centering     
         \includegraphics[width=0.49\textwidth]{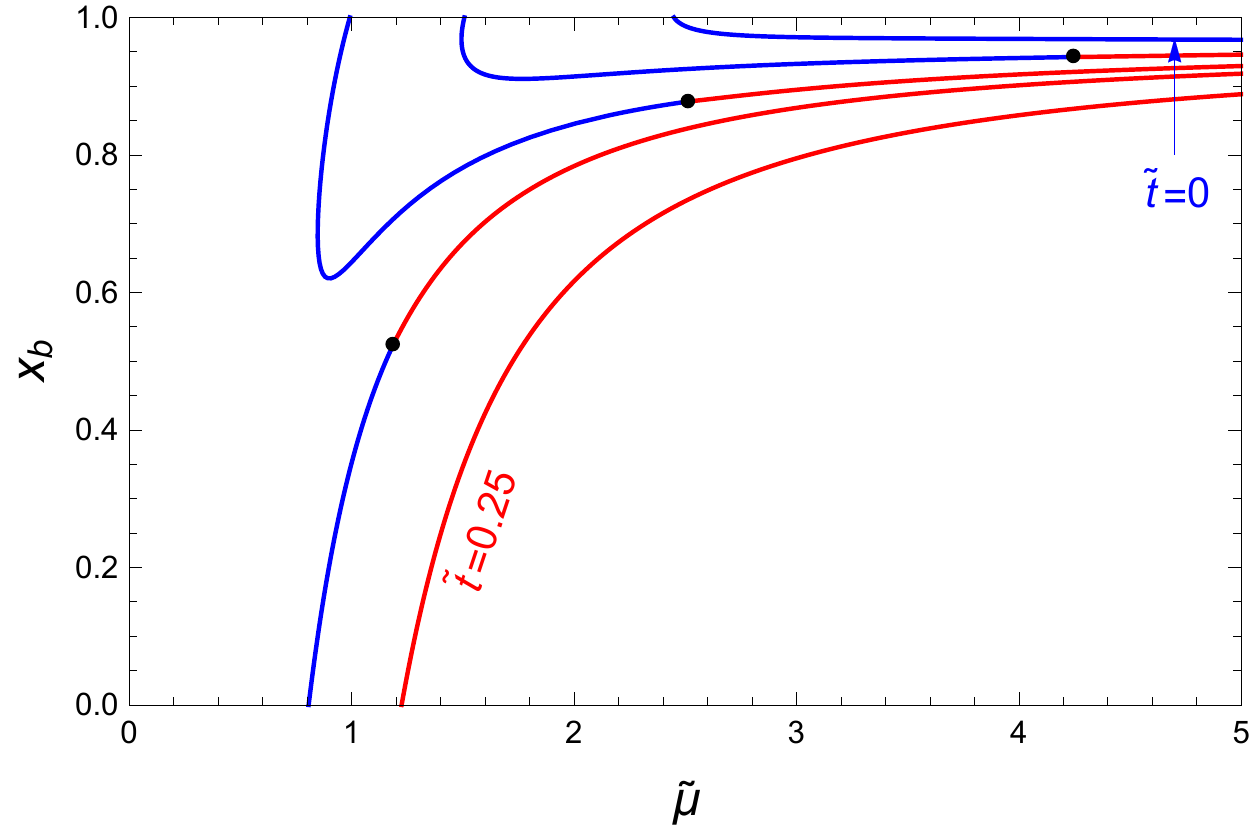}
         \includegraphics[width=0.49\textwidth]{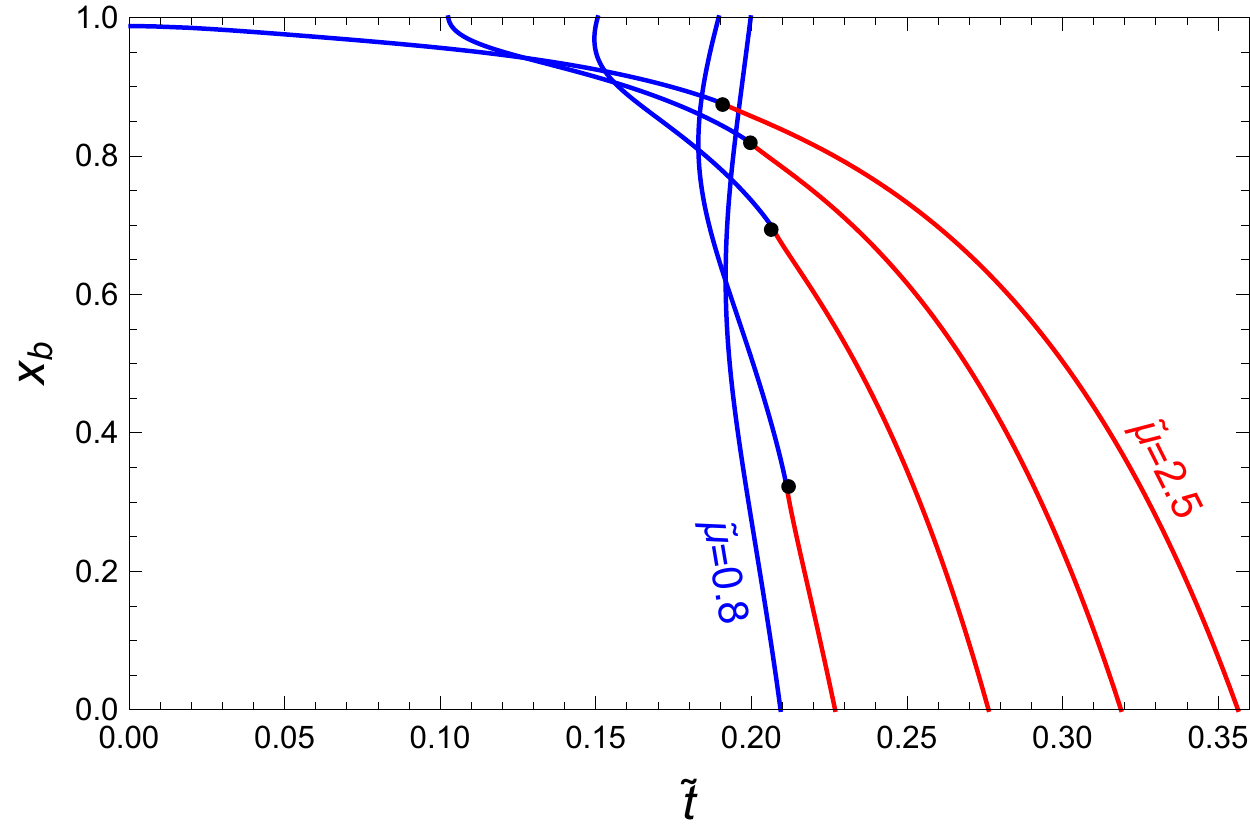}
         \caption{{\small As Fig.\ \ref{fig:xbchiral}, but for mass parameter $\tilde{\alpha}=0.2$, corresponding to the phase diagram in the right panel of Fig.\ \ref{fig:phasediagrams}. (Blue) segments to the left of the dots correspond to the LTQy phase, while (red) segments to the right of the dots correspond to the HTQy phase. Discontinuous phase transitions occur in the regions where the functions are two-valued (the exact location of the transitions is not indicated here). {\it Left panel:} $x_b$ as a function of chemical potential for temperatures $\tilde{t}= 0, 0.15, 0.19, 0.21, 0.25$. {\it Right panel:} $x_b$ as a function of temperature for chemical potentials $\tilde{\mu}=0.8, 1, 1.5, 2, 2.5$. }}
         \label{fig:Xb02}
\end{figure}

To understand the properties of the holographic quarkyonic phase, let us first  consider the baryon fraction $x_b$ (\ref{xbdef}), which is 0 in the quark phase, 1 in the baryonic phase, and $0<x_b<1$ in the quarkyonic phase. We present $x_b$ 
as a function of chemical potential and temperature in Figs.\ \ref{fig:xbchiral} and \ref{fig:Xb02}. Fig.\ \ref{fig:xbchiral} shows the baryon fraction in the chiral limit. In the left panel we  confirm the analytical result from Sec.\ \ref{sec:chiral}: at zero temperature, baryons account for a constant fraction of about 97\% of the baryon density in the quarkyonic phase. For larger temperatures, this fraction is approached from below as the chemical potential is increased. The plot also indicates the discontinuous baryonic-quarkyonic transition and, at sufficiently large temperatures, the continuous quark-quarkyonic transition. Stable quarkyonic matter in the chiral limit is always in the HTQy configuration (red curves in the figure). We have added the LTQy branch where it exists. This configuration is never stable in the chiral limit and even ceases to exist for large temperatures. 
The right panel of Fig.\ \ref{fig:xbchiral} shows the melting of baryons within the quarkyonic phase as the temperature is increased. As we have seen in the phase diagrams of Fig.\ \ref{fig:phasediagrams}, the transition from the quarkyonic to the pure quark phase is always continuous. This is confirmed in this plot, which shows that the baryon fraction goes to zero continuously.

The conclusions from Fig.\ \ref{fig:Xb02} are similar, but now, at nonzero pion mass, the LTQy configuration plays a prominent role. For instance, at zero temperature, the baryon fraction decreases  continuously from 1 to its asymptotic value, all within the LTQy configuration. For any nonzero, but not too large, temperature, there is a transition from the LTQy to the HTQy configuration at some chemical potential. At sufficiently large temperatures, the quarkyonic phase appears through a continuous transition into the HTQy configuration, as already seen in the chiral limit. Both panels of Fig.\ \ref{fig:Xb02} show regions where $x_b$ is a two-valued function of either temperature or chemical potential. In these cases, there is a jump in the baryon fraction, whose precise location cannot be read off of these curves. The determination of the critical point requires the calculation of the free energy, which we have done for the phase diagrams in Fig.\ \ref{fig:phasediagrams}, but we have not indicated the phase transition in Fig.\ \ref{fig:Xb02} in order to keep the plots simple. 

\begin{figure}[t]
       \centering
       \includegraphics[width=\textwidth]{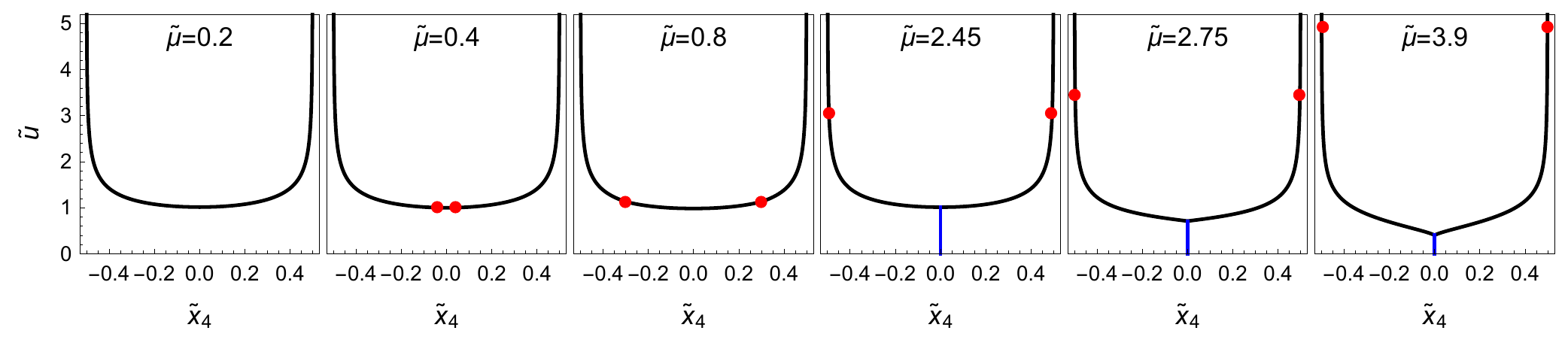}
       
       \vspace{-0.4cm}
       \includegraphics[width=\textwidth]{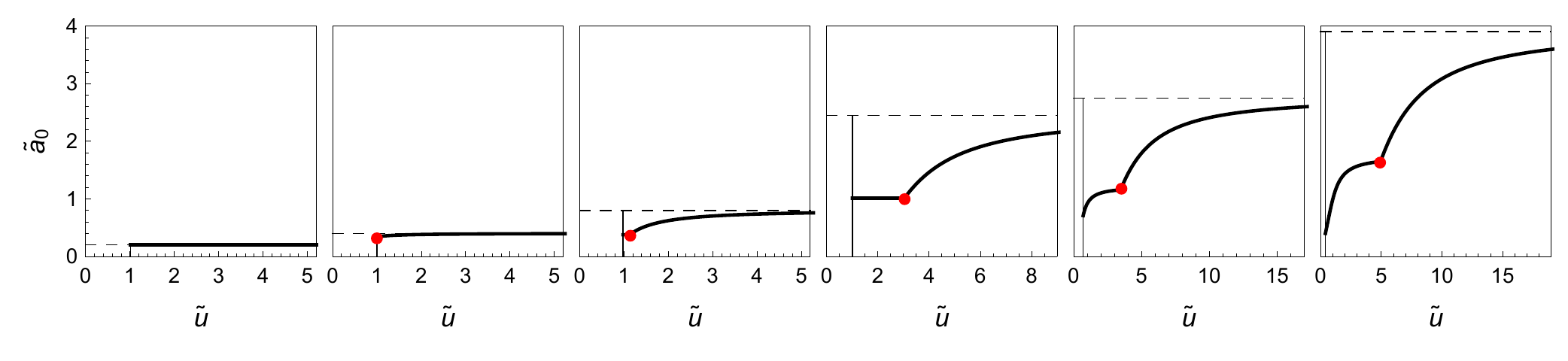}
       
       \vspace{0.5cm}
       \includegraphics[width=\textwidth]{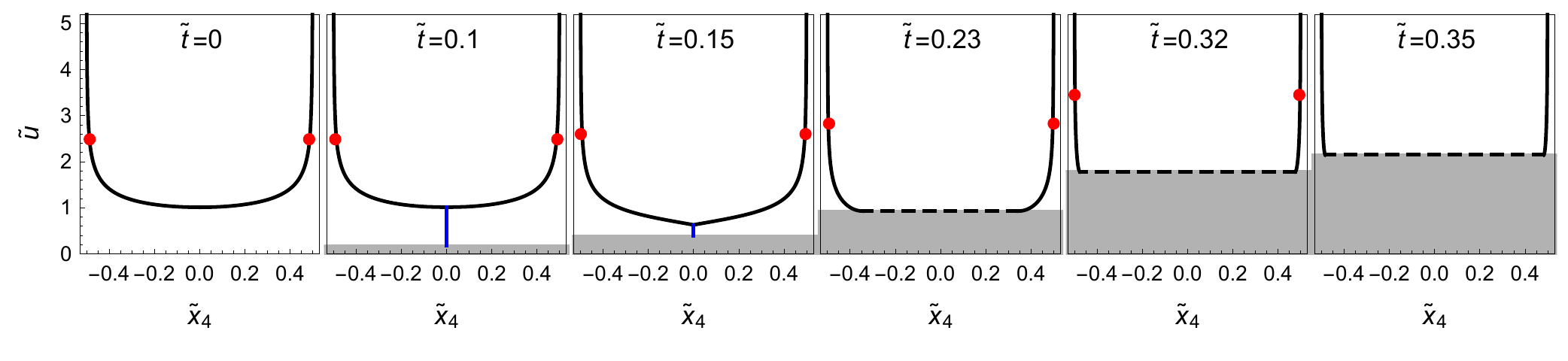}
         
      \vspace{-0.2cm}
      \includegraphics[width=\textwidth]{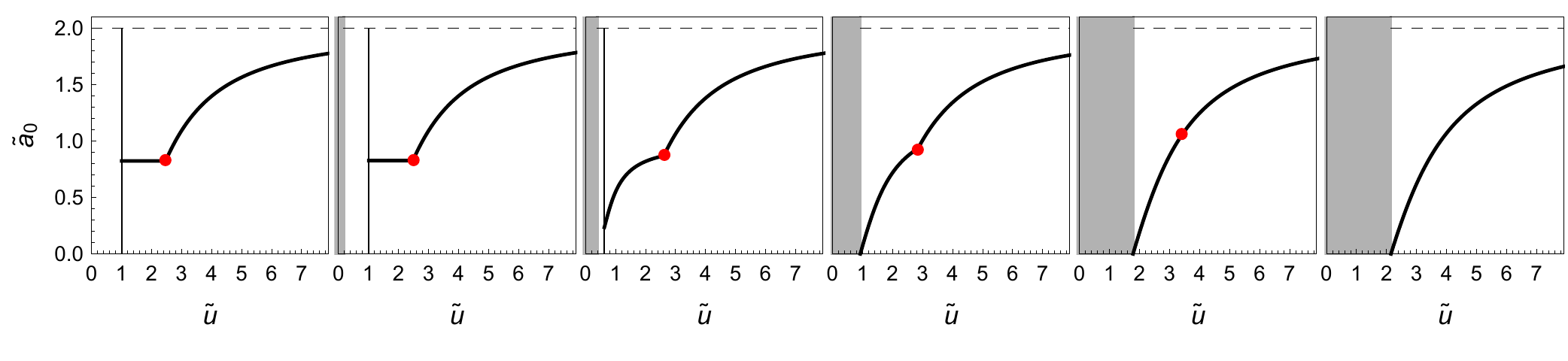}
        \caption{{\small Embedding of the flavor branes in the subspace spanned by the holographic coordinate $\tilde{u}$ and the compactified direction $\tilde{x}_4$ together with the corresponding abelian gauge field $\tilde{a}_0(\tilde{u})$ on (one half of) the branes  for zero temperature and various chemical potentials (upper two rows) and a fixed chemical potential $\tilde{\mu}=2$ and various temperatures (lower two rows). For all panels, the mass parameter is $\tilde{\alpha}=0.2$, as in the right panel of Fig.\ \ref{fig:phasediagrams}. The (red) dots indicate the location of the pointlike instantons $\tilde{u}_b$, the (blue) straight lines indicate strings from the horizon $\tilde{u}_T$ to the tip of the connected branes $\tilde{u}_c$. The region $\tilde{u}<\tilde{u}_T$ is shaded to indicate the location of the horizon. In the plots of $\tilde{a}_0$, the thin vertical lines mark $\tilde{u}_c$, and the horizontal dashed lines indicate $\tilde{\mu}$, to which the gauge field asymptotes. Phases where instantons and strings coexist or where instantons exist together with branes that reach the horizon are quarkyonic.}}
         \label{fig:EmbT}
\end{figure}

\subsection{Holographic picture of quarkyonic matter}
\label{sec:embed}

We further illustrate our results, and make the connection to the geometry of the model, by computing the embedding of the flavor branes
$x_4(u)$ and the abelian gauge field $\hat{a}_0(u)$. The results are shown in Fig.\ \ref{fig:EmbT} for zero temperature, $\tilde{t}=0$, and various chemical potentials (upper two rows) and for a fixed chemical potential, $\tilde{\mu} = 2$, and various temperatures (lower two rows). We have chosen to present the results for the "heavy" case $\tilde{\alpha}=0.2$, and the figure only shows the energetically preferred configuration for any given $\tilde{t}$ and $\tilde{\mu}$. 

It is instructive to walk through the plots step by step. Let us begin with the first two rows, i.e., zero temperature. We start from the mesonic phase, where the embedding does not depend on $\mu$, and the gauge field is constant. As we increase the chemical potential, pointlike instantons appear at the tip of the connected flavor branes. This occurs at the point given by the condition (\ref{nbonset}), with infinitesimally small density. Then, immediately two baryon layers  move up in the holographic direction. Due to the flatness of the embedding around its tip, the location $u_b$ stays almost constant for small densities although the baryon layers visibly increase their distance. As the baryons move in the holographic direction towards larger values of $u$ the density increases. At some point, here at $\tilde{\mu}\simeq 2.45$, given by the condition (\ref{nqonset}), it becomes favorable to add strings. For the given parameters, the onset of strings is continuous, i.e., the quark contribution to the baryon density is infinitesimally small at that point. We have indicated the strings by a (blue) line from the horizon to the tip of the connected branes. The fact that we start off with infinitesimally small quark density is also reflected by the shape of the embedding at the tip, which is smooth and only for larger chemical potentials develops a cusp that is clearly visible in the figure. The embedding also has a cusp at the location of the baryon layers, but this cusp only becomes visible by naked eye on a smaller scale. In contrast, the cusp of the gauge field at the location of the instantons is clearly visible. The profile of the gauge field changes qualitatively when strings are added: while the profile is constant for $u<u_b$ in the purely baryonic phase, it develops a gradient in this domain due to the string sources. Once the quarkyonic phase has become favored, the baryons keep moving towards the holographic boundary while the branes themselves move towards the horizon, approaching the scenario of straight branes for asymptotically large chemical potentials. 

In the lower two rows of the figure we walk, at a given, rather large, chemical potential from the baryonic phase at zero temperature to the pure quark phase at large temperatures via  the quarkyonic phase in between. Again, we show the continuous onset of quarkyonic matter, in this case at $\tilde{t}\simeq 0.1$, indicated by the strings and a smooth embedding. Now, with increasing temperature, the horizon "catches up" with the branes, such that the system transitions from the LTQy phase into the HTQy phase. The plots confirm our earlier observation that in the case of a nonzero pion mass the branes reach the horizon tangentially. This allows for the branes to connect smoothly along the horizon, which we have indicated by a dashed line. Along the dashed line, the gauge field vanishes, due to the boundary condition $\hat{a}_0(u_T)=0$, and this segment is irrelevant for calculation of the free energy. As we keep increasing the temperature, the location of the baryons $u_b$ keeps increasing slightly, although the distance to the horizon $u_b-u_T$ decreases. 
Eventually, the baryon fraction in the quarkyonic phase goes to zero. At $\tilde{t}\simeq 0.32$, there is a continuous transition to the pure quark phase, i.e., the system decides to remove all instantons from the flavor branes. In the profiles of the gauge fields this is manifest in a less pronounced cusp which eventually is smoothed out completely. Again, as for large chemical potentials, the embedding of the flavor branes becomes straight for asymptotically large temperatures. 

With the help of this figure we may further refine our holographic picture of quarkyonic matter. As the solutions of the equations of motion (\ref{EOM}) show,  the embedding of the flavor branes and the behavior of the abelian gauge field for $u<u_b$ is not affected by the presence of the baryons. Therefore, in the infrared regime, up to the (red) dots in the figure, the physics of the quarkyonic phase is entirely determined by quarks. (It can be expected that for nonzero-width instantons this separation of regimes is less sharp.) Interestingly, we observe that $\hat{a}_0(u)$ becomes flat as it approaches $u_b$, i.e., the gauge field behaves in the infrared as if there was a second chemical potential $\hat{a}_0(u_b)$ to which $\hat{a}_0$ asymptotes as we approach the location of the instantons, just like $\hat{a}_0$ asymptotes to the actual chemical potential as $u\to \infty$.  This is reminiscent of the layered structure in momentum space in the weak-coupling picture of the quarkyonic phase. And just like a Fermi momentum, $u_b$ moves towards higher energies as the chemical potential is increased, while increasing the temperature at fixed chemical potential only leads to a very small change in the location of the baryons. A similar relation between the baryon distribution in the bulk and the Fermi surface was already suggested in the literature, albeit in the absence of quark sources  \cite{Rozali:2007rx,Elliot-Ripley:2016uwb}. The ultraviolet regime, $u>u_b$, however, is affected by both quarks and baryons. This goes beyond the simple weak-coupling picture of a purely baryonic layer in momentum space on top of the quark Fermi sea.

\subsection{Quark-hadron continuity}
\label{sec:continuity}

As we have seen in the discussion of the phase diagrams in Fig.\ \ref{fig:phasediagrams}, by tuning the  value of the pion mass we find a regime where we can go continuously from baryonic matter via quarkyonic matter to pure quark matter. This continuity requires us to go to nonzero temperature, while at zero temperature we have a baryonic-quarkyonic continuity, with  pure quark matter never being realized in the given approximation. Also, the continuity is not a completely smooth crossover, it rather involves second-order transitions, due to our approximations and presumably also due to the large-$N_c$ regime to which the evaluation of our holographic model is constrained.
Nevertheless, this feature of the model is remarkable since it may help to shed some light on the question of whether there is a zero-temperature quark-hadron continuity in QCD, or whether there is (at least) one discontinuous transition, as predicted by many phenomenological models, which however usually contain either quark or nucleon degrees of freedom, but not both (for phenomenological models where quarks and nucleons are put together see for instance Refs.\ \cite{Dexheimer:2009hi,Aryal:2020ocm,Marczenko:2020jma}). Our picture of a possible continuity is of course simplified due to the lack of Cooper pairing. Superfluidity and superconductivity may induce further, possibly discontinuous, phase transitions, although it has been argued that even in the presence of color superconductivity, quarkyonic matter may enable a continuous transition from nuclear to quark matter \cite{Fukushima:2015bda}. 

\begin{figure}[t]
    \centering
         \includegraphics[width=0.48\textwidth]{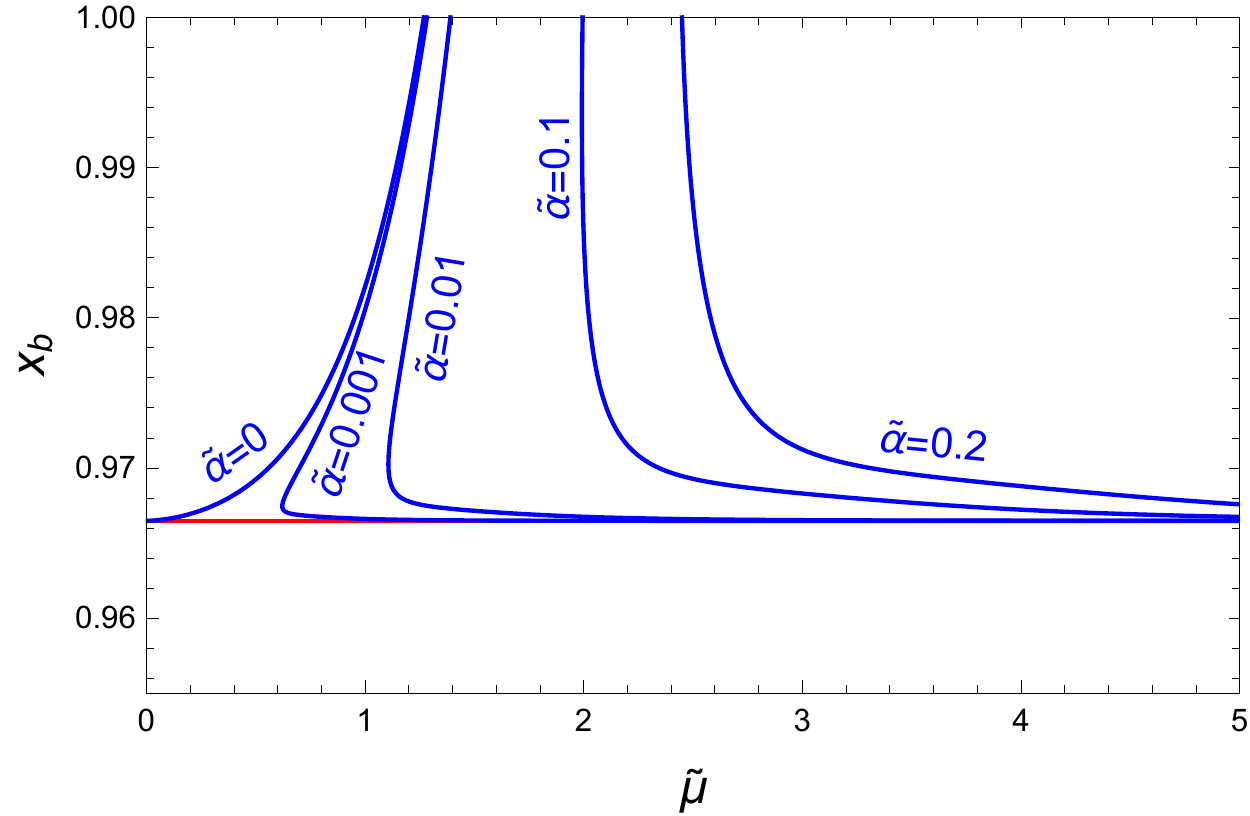}
         \caption{{\small Zero-temperature baryon fraction $x_b$ in the quarkyonic LTQy configuration 
         as a function of the chemical potential for different values of the mass parameter $\tilde{\alpha}$. For nonzero and not too large $\tilde{\alpha}$ there is a discontinuous transition from the baryonic phase (where $x_b=1$) to the quarkyonic LTQy phase, indicated by the two-valuedness of the curves. This transition  becomes continuous at around $\tilde{\alpha}\simeq 0.1$. In the chiral limit, $\tilde{\alpha}=0$, the entire LTQy branch is unstable, and  there is a discontinuous transition from baryonic matter to HTQy quarkyonic matter, which has a constant baryon fraction $x_b\simeq 0.966471$, shown by the (red) horizontal line. }}
         \label{fig:xmu}
\end{figure}

In this section, we will restrict ourselves to zero temperature, i.e., we actually focus on the  baryonic-quarkyonic continuity. To this end, we first come back to the baryon fraction $x_b$, which is plotted in Fig.\ \ref{fig:xmu} as a function of the chemical potential for different values of the quark mass parameter. This plot shows how the discontinuous transition at small $\tilde{\alpha}$ turns into a continuous transition at around $\tilde{\alpha}\simeq 0.1$. Using the fit discussed below Eq.\ (\ref{mpi}), this corresponds to a pion mass of about $m_\pi\simeq 980\, {\rm MeV}$. 

\begin{figure}[t]
    \centering
       \hspace{1cm} \includegraphics[width=0.35\textwidth]{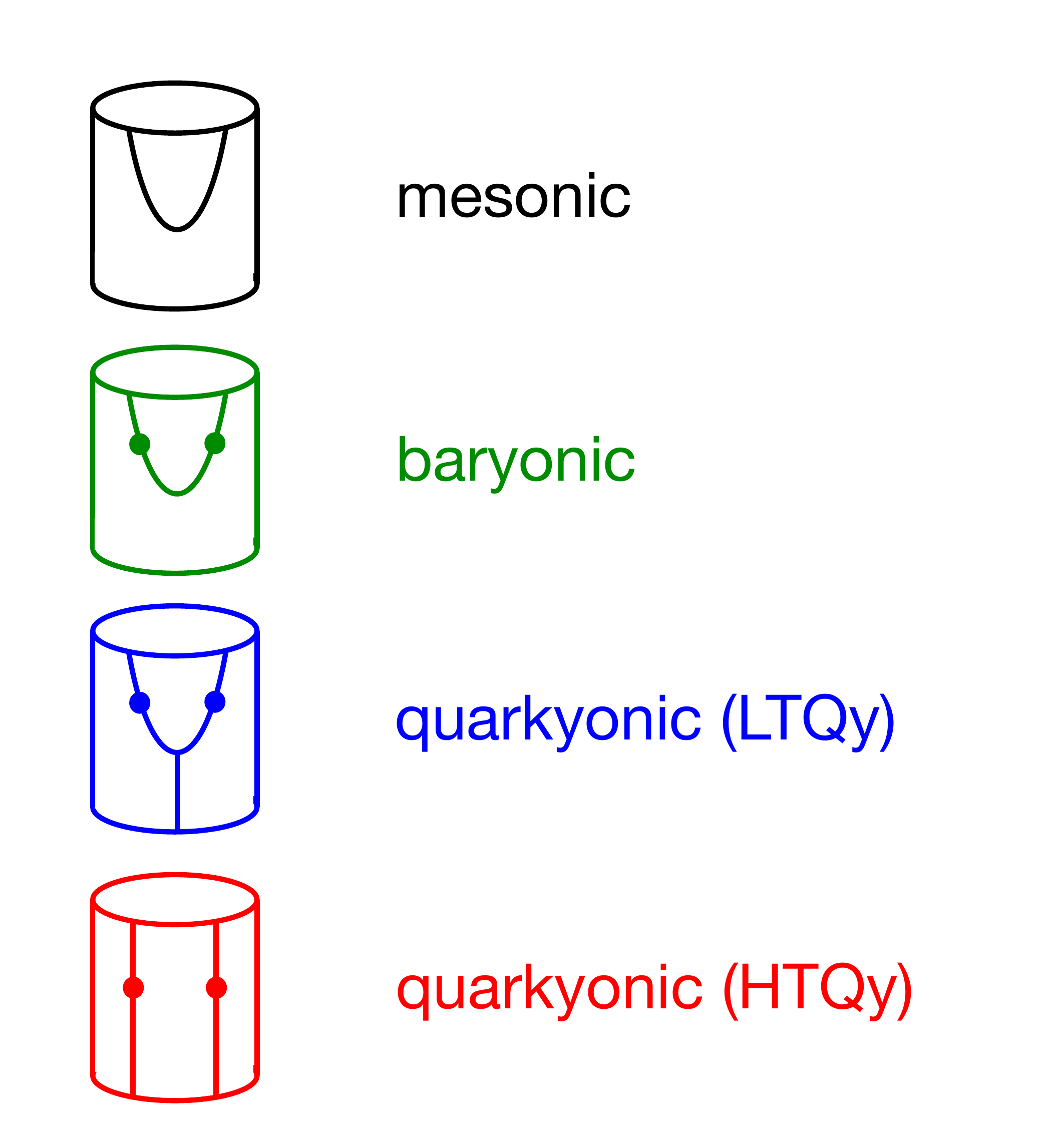}\hspace{1cm}
        \includegraphics[width=0.49\textwidth]{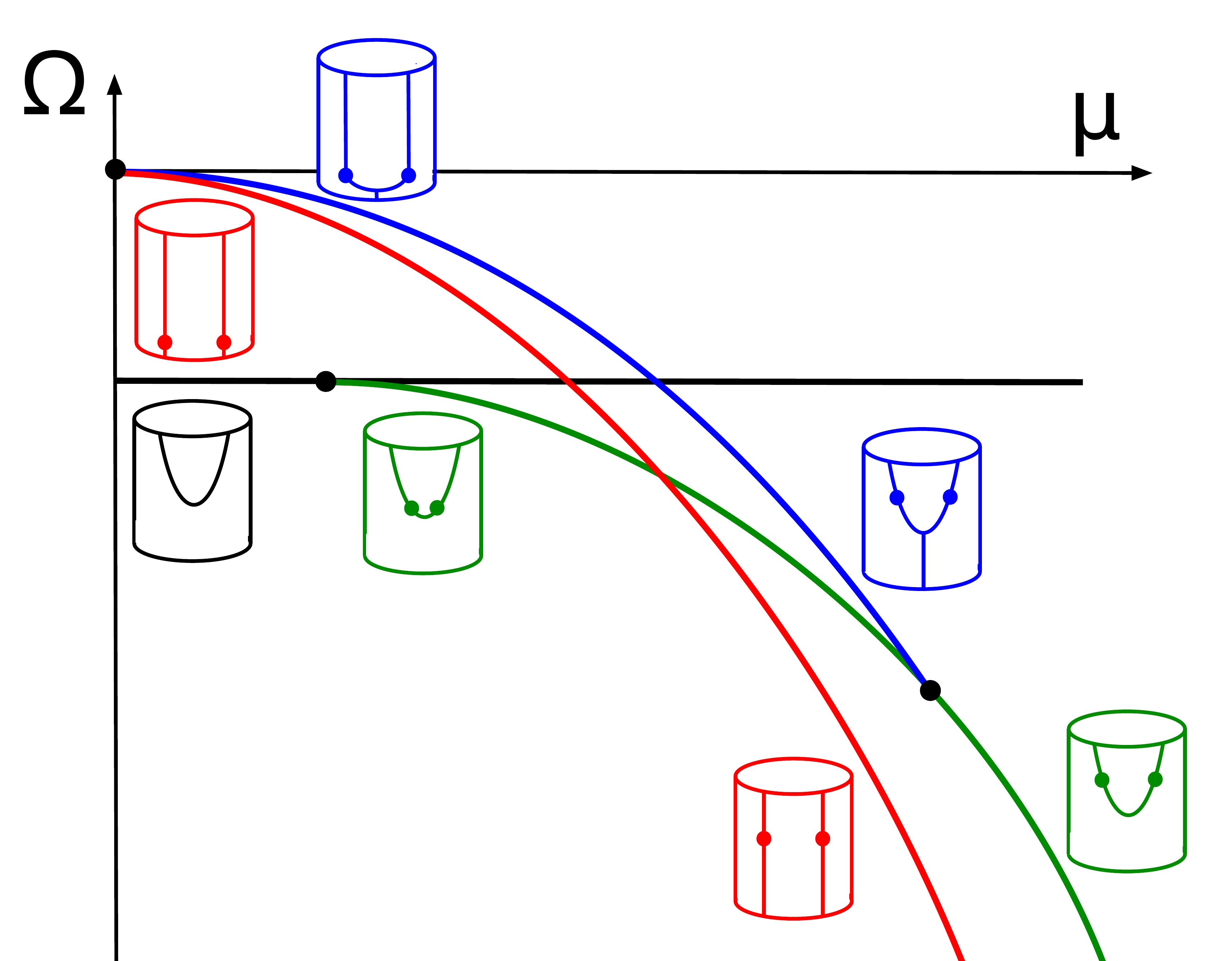}
        \includegraphics[width=0.49\textwidth]{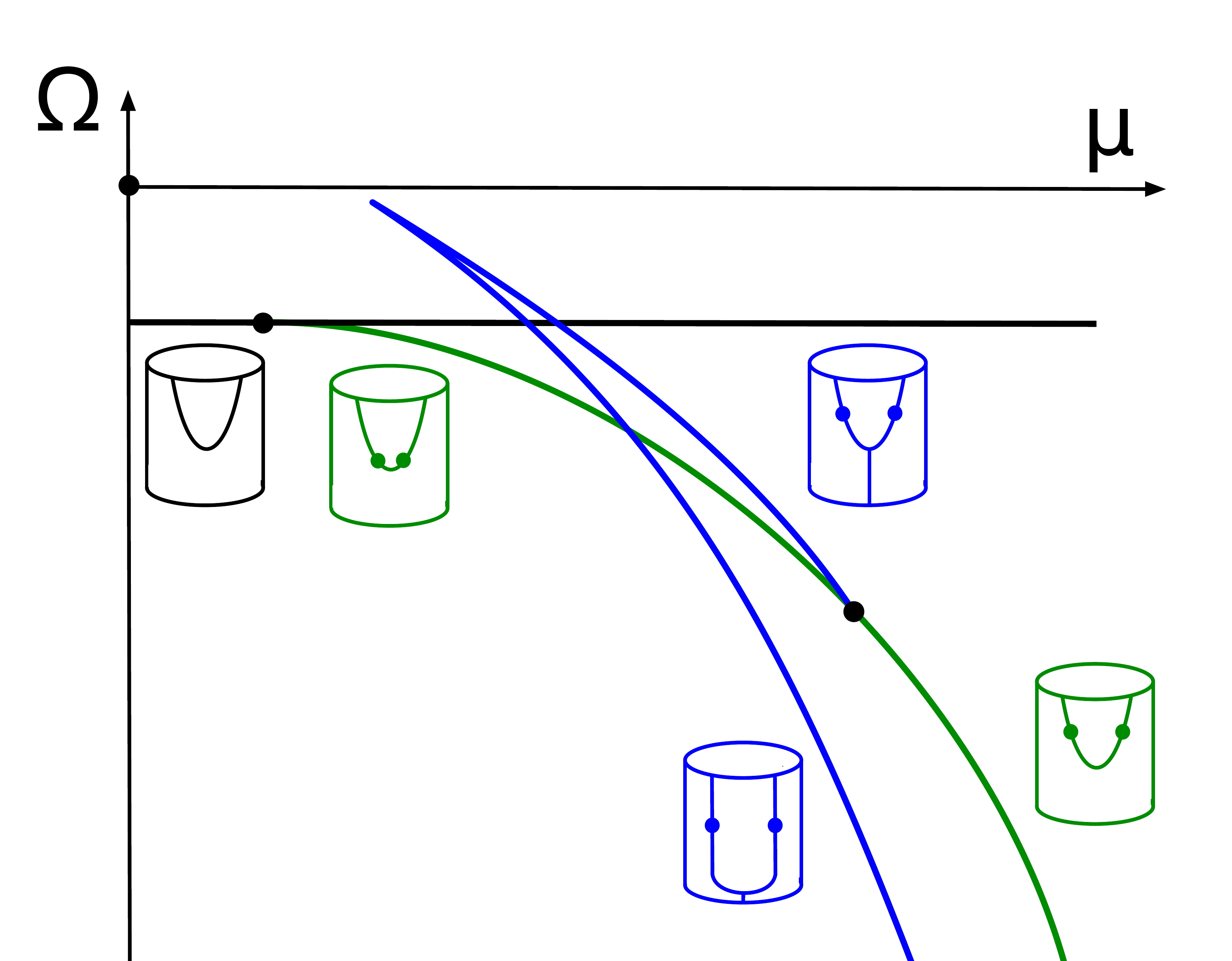}\hspace{-0.3cm}
        \includegraphics[width=0.49\textwidth]{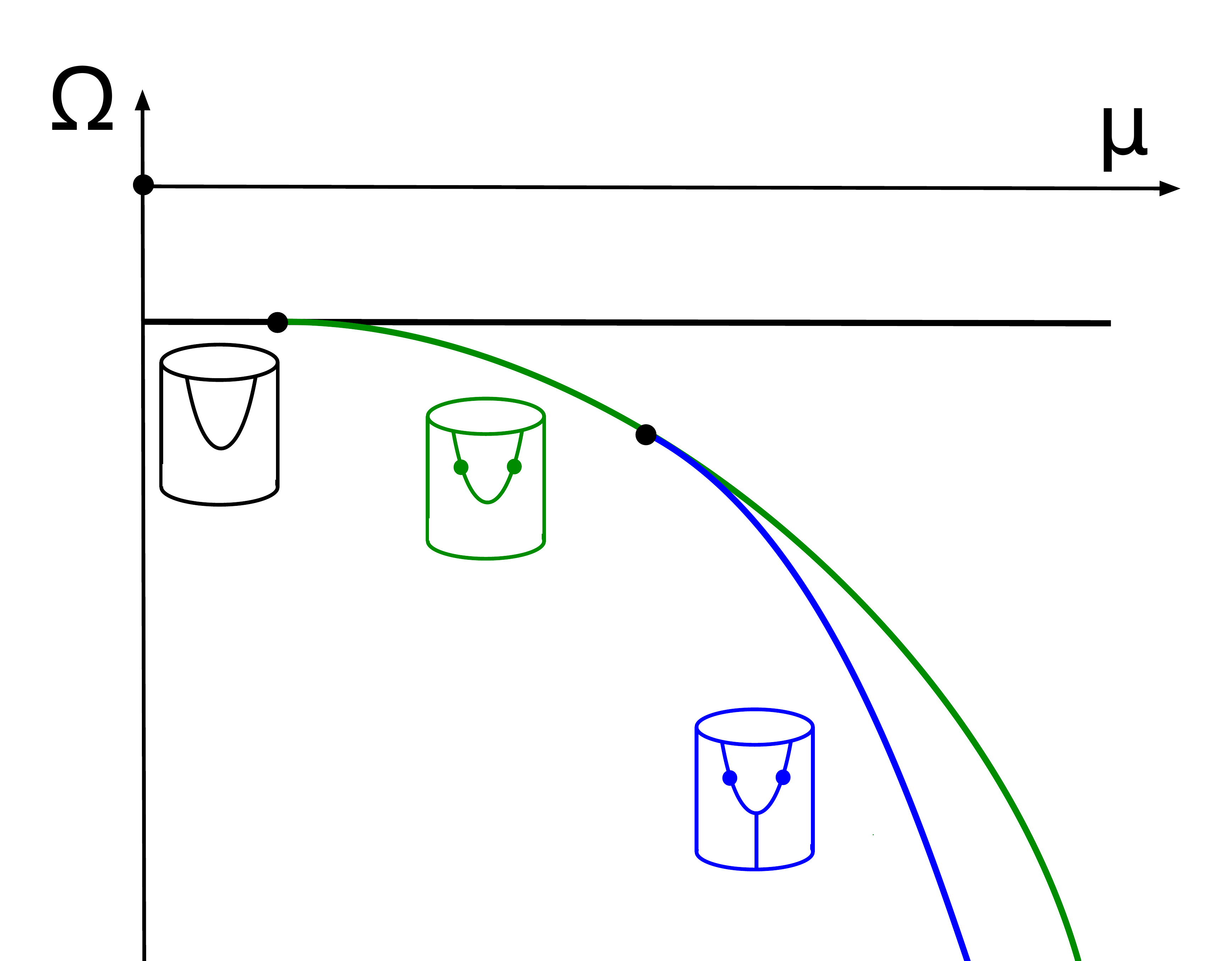}
         \caption{{\small Schematic plots of the free energy density $\Omega$ at its local extrema as a function of the chemical potential $\mu$ for zero (upper right), intermediate (lower left) and large (lower right) pion mass. The figure illustrates the appearance of a continuity between baryonic and quarkyonic matter as the pion mass is increased, and shows that this continuity can already be anticipated from the upper right plot due to the existence of the unstable LTQy branch.}}
         \label{fig:schematicQH}
\end{figure}

The continuity is illustrated further in Fig.\ \ref{fig:schematicQH}. This figure is similar to Fig.\ 3 in Ref.\ \cite{BitaghsirFadafan:2018uzs}, where it was already conjectured that a quark-hadron continuity is conceivable in the Witten-Sakai-Sugimoto model at nonzero current quark mass. Here we have found a realization of the continuity by allowing for quarkyonic matter, albeit for unrealistically large pion masses. It remains to be seen whether and how this conclusion is changed using the more realistic baryons of Ref.\ \cite{BitaghsirFadafan:2018uzs}. Our Fig.\ \ref{fig:schematicQH}
schematically shows the free energy density as a function of chemical potential for zero (upper right), intermediate (lower left) and large (lower right) pion mass, together with the corresponding brane embeddings. We have chosen a schematic representation for clarity. Had we plotted the actual results for the free energies, their differences would have been very difficult to resolve, although of course Fig.\ \ref{fig:schematicQH} is in qualitative agreement with our numerical results. Besides the stable states, the figure includes local extrema of the free energy density which are not global minima. In particular, some of the branches are local maxima at the boundary of our multi-parameter space, which are not stationary: as discussed below Eq.\ (\ref{eqA}), mesonic and baryonic phases have branches where switching on a density contribution immediately lowers the free energy, corresponding to a $\land$-shaped local maximum if negative densities are taken into account. In the figure, this concerns the (black) mesonic branch on the right-hand side of the baryon onset and the (green) baryonic branch on the right-hand side of the quarkyonic onset.  

In the 
upper right plot, which shows the chiral limit, at first sight, the (green) baryonic and (red)  quarkyonic curves appear unrelated. However, they turn out to be continuously connected upon including the (blue) LTQy configuration (the points on this curve correspond to local maxima as well, but stationary). This is very similar to Ref.\  \cite{BitaghsirFadafan:2018uzs}, where the connection between pure baryonic and pure quark phases was pointed out. As  the current quark mass is set to zero, quarks may in principle appear at arbitrarily small chemical potentials, although this possibility is of course energetically disfavored. Moreover, since the dimensionless baryon mass is $u_b/3$ at zero temperature, see Eq.\ (\ref{baryonmass}), baryons can also become arbitrarily light. Therefore, the (red) quarkyonic branch -- given by Eq.\ (\ref{OmMix00}) --  reaches all the way back to zero chemical potential, and the connection between baryonic and quarkyonic phases can only be made through the origin of the diagram. 

As expected, and as anticipated in Ref.\ \cite{BitaghsirFadafan:2018uzs}, the branches move away from the origin as the current quark mass is switched on. 
In this case, the straight brane solution ceases to exist, hence the absence of a (red) HTQy curve. 
Instead, the quarkyonic LTQy solution turns around  before reaching the origin and becomes favored after crossing the pure baryonic line. Now, for sufficiently large masses ($\tilde{\alpha}\gtrsim 0.1$, as Fig.\ \ref{fig:xmu} has shown), the quarkyonic solution becomes single-valued, resulting in a transition where the free energy and its first derivative are continuous. We have thus obtained a concrete holographic realization of a zero-temperature baryonic-quarkyonic continuity. Interestingly, as our phase diagram shows, this continuous transition turns into a discontinuous one at larger temperatures, giving rise to a high-density critical point, not unlike the one predicted in Ref.\ \cite{Hatsuda:2006ps}.

\begin{figure}[t]
    \centering
        \includegraphics[width=0.5\textwidth]{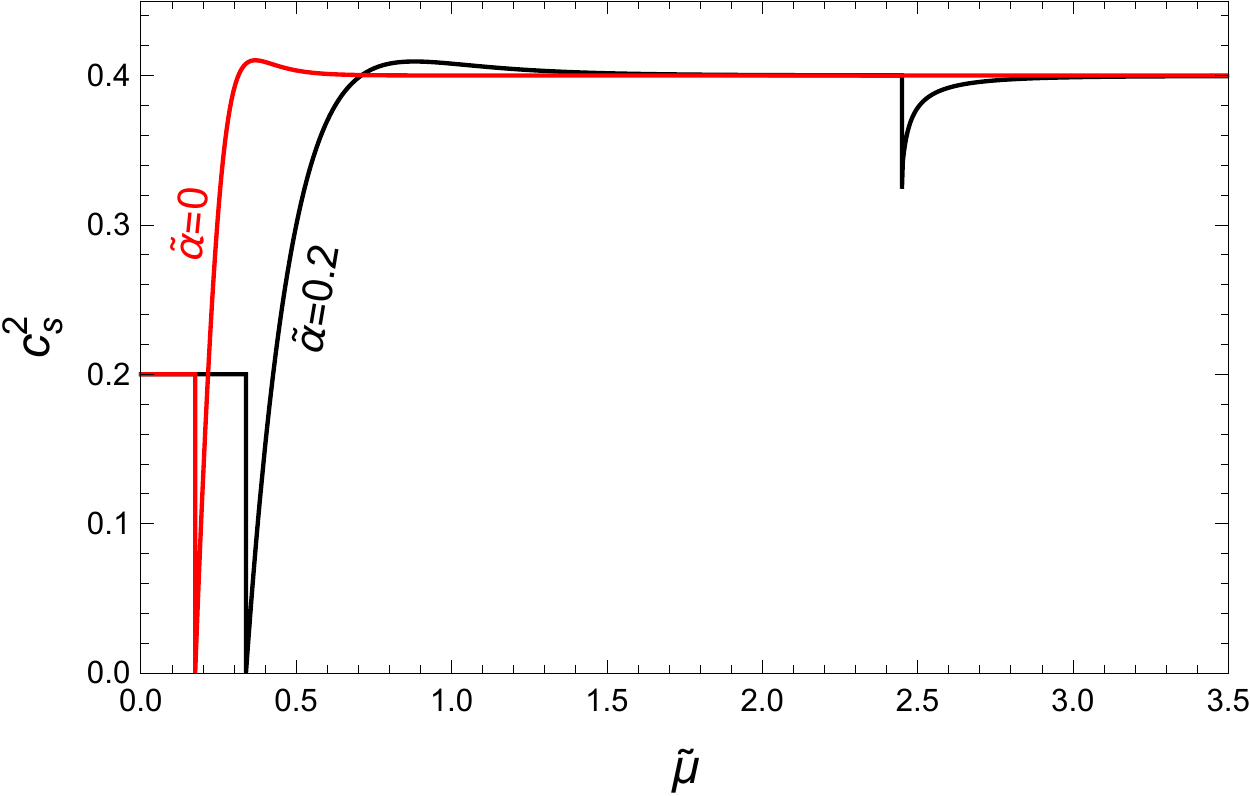}
         \caption{{\small Speed of sound squared at zero temperature as a function of the chemical potential for two different values of the mass parameter $\tilde{\alpha}$. Both curves have a discontinuity at the transition from mesonic to baryonic matter at small $\tilde{\mu}$ and at the transition from baryonic to quarkyonic matter at large $\tilde{\mu}$ (the latter discontinuity in the chiral limit $\tilde{\alpha}=0$ occurs at $\tilde{\mu}\simeq 1.03$ and is too small to be visible here). }}
         \label{fig:Cs2}
\end{figure}

By computing the speed of sound $c_s$ we can further analyze the nature of the transition because the calculation involves second derivatives of the free energy. Since in this section we are only interested in zero temperature, we may use the 
definition 
\begin{equation}
 c_s^2  
 =\frac{n}{\mu} \left(\frac{\der n}{\der \mu}\right)^{-1}\, . \label{cs}   
\end{equation}
(The mesonic phase has zero baryon density, and thus Eq.\ (\ref{cs}) cannot be used, and we have to compute the zero-temperature limit following Appendix A of Ref.\ \cite{Kovensky:2019bih}.) 
We plot the result of the stable phases for the two different mass parameters $\tilde{\alpha}=0$ and $\tilde{\alpha}=0.2$ as a function of the chemical potential in Fig.~\ref{fig:Cs2}. Although less relevant for our present purpose, we see that there is a large jump at the second-order baryon onset. More importantly, we observe a relatively large discontinuity in the speed of sound in the "heavy" scenario at the baryonic-quarkyonic transition. This discontinuity also exists in the chiral limit at the transition point, but it is too small to be seen on the given scale. Therefore, somewhat surprisingly, the jump in the speed of sound is {\it smaller} in the case of a first-order transition (where the density jumps) than in the case of a second-order transition (where the density does not jump). Mathematically, this is easily understood: it simply means that in one case we have a discontinuous function, namely the density, whose derivatives on both sides of the discontinuity are almost identical, while in the other case we have a continuous function with a (relatively strong) cusp.  

We also see that the speed of sound is non-monotonic in the baryonic phase. This was already observed in the same model with more realistic baryons, where this non-monotonicity was even more pronounced, with a larger maximum of the speed of sound   \cite{BitaghsirFadafan:2018uzs}. In the quarkyonic phase, the zero-temperature speed of sound is constant in the chiral limit, $c_s^2 = 2/5$. This follows from the free energy density (\ref{OmMix00}) and the definition (\ref{cs}). The same result holds for the pure quark phase
\cite{Kulaxizi:2008jx,BitaghsirFadafan:2018uzs}, which is only metastable at zero temperature, as already mentioned below Eq.\ (\ref{OmHTQ00}). In contrast, at a nonzero pion mass we see that the speed of sound in the quarkyonic phase increases monotonically and approaches $c_s^2 = 2/5$ at large chemical potentials. 
This is different from QCD, where the speed of sound approaches the conformal limit $c_s^2 = 1/3$ asymptotically. 
The behavior of the speed of sound is also interesting in comparison to the model for quarkyonic matter of Ref.\ \cite{McLerran:2018hbz}. In that model,  the speed of sound is non-monotonic in the quarkyonic phase. This feature was highlighted since astrophysical observations, most notably the masses of the heaviest neutron stars, together with asymptotic freedom of QCD, suggests that the speed of sound in QCD is indeed non-monotonic. Our results suggest that the quarkyonic speed of sound does not necessarily have this feature, and instead ordinary baryonic matter may account for the non-monotonicity.

\section{Summary and outlook}

We have discussed new configurations in the holographic Witten-Sakai-Sugimoto model and pointed out their significance for the 
phase diagram. Working in the deconfined geometry and the decompactified limit of the model, we have found that baryons can be added to the previously discussed low-temperature quark (LTQ) and high-temperature quark (HTQ) configurations. We have interpreted these new configurations as 
holographic realizations of quarkyonic matter, a low-temperature version (LTQy), containing strings stretching from the horizon to the tip of the connected flavor branes, and a high-temperature version (HTQy), where the flavor branes reach the horizon. In both cases, two layers of pointlike instantons are placed at dynamically determined points on the flavor branes. In a way, these geometries can be viewed as a strong-coupling version of the weak-coupling picture of quarkyonic matter based on a quark Fermi sea enclosed by a baryonic layer in momentum space. 

We have shown that holographic quarkyonic matter covers a significant part of the phase diagram. For low temperatures, as the density is increased, there is a discontinuous phase transition from purely baryonic matter to quarkyonic matter. This phase transition coincides with the chiral phase transition, i.e., our model predicts quarkyonic matter to be (approximately) chirally symmetric. At zero temperature, we have found that  baryons contribute a large fraction to the total baryon density in the quarkyonic phase, 97\% in the chiral limit. At the same time, the quarkyonic phase can behave in a quark-like manner, as we demonstrated with the speed of sound. In the approximation used here, quarkyonic matter does not transition to pure quark matter at ultra-high densities. However, baryons in the quarkyonic phase do melt as the temperature is increased, resulting in a continuous high-temperature phase transition to pure quark matter. 

Furthermore, we have explored the changes in the phase diagram under variation of the pion mass. Most notably, we found that for unrealistically large pion masses, $m_\pi\gtrsim 1\, {\rm GeV}$, the baryonic-quarkyonic transition becomes continuous for low temperatures, resulting in a high-density critical point in the phase diagram. 

We have made use of various approximations that can be improved in the future. Most importantly, our baryons are delta-peaks rather than instantons with nonzero widths. We have resorted to this approximation to simplify the study of the quarkyonic configurations, and we have largely ignored the possibility of a multi-layer structure of the baryons. More sophisticated approximations of holographic baryonic matter exist in the literature \cite{Ghoroku:2012am,Li:2015uea,Preis:2016fsp,BitaghsirFadafan:2018uzs}, and it would be interesting to implement these improvements in the study of holographic quarkyonic matter. We have ignored any form of superconductivity, which is expected in high-density QCD \cite{Alford:2007xm} and for which simple holographic realizations exist \cite{BitaghsirFadafan:2018iqr}. We have also ignored any inhomogeneous structures such as chiral spirals, which are predicted to occur in quarkyonic matter \cite{Kojo:2009ha,Kojo:2011cn} and which have been discussed in the presence of a magnetic field in the Witten-Sakai-Sugimoto model \cite{Rebhan:2008ur,BallonBayona:2012wx}. 
Our approach has been restricted to the deconfined geometry of the model and we have ignored the confined geometry, which is preferred at small temperatures. Finding a solution that describes quarkyonic matter in the confined geometry would be desirable, but is probably challenging since our construction cannot be straightforwardly implemented in the absence of a horizon. Finally, we have 
restricted ourselves to the quenched $N_f \ll N_c$ regime.  
It would be interesting to explore the consequences of including the backreaction of the flavor branes on the background geometry, possibly along the lines of Refs.\ \cite{Bigazzi:2014qsa,Li:2016gtz}.

Since our quarkyonic matter is chirally symmetric, a calculation of the hadron spectrum within this phase would be interesting. As in different approaches dealing with chiral symmetry restoration in a confined phase  \cite{Wagenbrunn:2007ie,Glozman:2007tv,Glozman:2012fj} we would expect chiral multiplets, in contrast to non-degenerate spectra in the chirally broken purely baryonic phase. It would also be interesting to use and extend our present results for phenomenological applications. For instance, one might ask whether quarkyonic matter exists in the interior of neutron stars or whether it might be created in a neutron star merger. This question can be addressed by computing the equation of state and the resulting mass-radius relations and tidal deformability, as recently done within the same model, but without any form of quark matter \cite{Hirayama:2019vod}, and in other holographic approaches  \cite{Hoyos:2016zke,Chesler:2019osn,Ecker:2019xrw,Fadafa:2019euu,Jokela:2020piw}. We have pointed out that in the given approximation the zero-temperature transition to quarkyonic matter occurs at densities probably too large to be reached in neutron stars, although at nonzero temperatures the transition density becomes smaller and is possibly reached in a merger
process. In any case, improving the approximation is likely to lead to quantitative changes, and previous studies suggest that a more refined treatment of the instantons moves the chiral transition towards lower densities \cite{Li:2015uea}. For a discussion of realistic neutron star matter it would also be important to include a nonzero isospin chemical potential into the calculation and allow for non-degenerate quark masses.

\acknowledgments

We would like to thank K.\ Bitaghsir Fadafan, N.\ Evans, M.\ J\"{a}rvinen, and R.\ Pisarski for valuable comments and discussions. This work is supported by the Leverhulme Trust under grant no RPG-2018-153. The work of N.K.\ has additionally been supported  by the National Agency for the Promotion of Science and Technology of Argentina (ANPCyT-FONCyT) Grant PICT-2017-1647. A.S.\ is supported by the Science \& Technology Facilities Council (STFC) in the form of an Ernest Rutherford Fellowship.  


\bibliographystyle{JHEP}
\bibliography{refs}

\providecommand{\href}[2]{#2}\begingroup\raggedright\begin{thebibliography}{10}

\bibitem{Alford:2007xm}
M.~G. Alford, A.~Schmitt, K.~Rajagopal, and T.~Sch{\"a}fer, {\it {Color
  superconductivity in dense quark matter}},  {\em Rev. Mod. Phys.} {\bf 80}
  (2008) 1455--1515, [\href{http://arxiv.org/abs/0709.4635}{{\tt
  arXiv:0709.4635}}].

\bibitem{McLerran:2007qj}
L.~McLerran and R.~D. Pisarski, {\it {Phases of cold, dense quarks at large
  $N_c$}},  {\em Nucl.Phys.} {\bf A796} (2007) 83--100,
  [\href{http://arxiv.org/abs/0706.2191}{{\tt arXiv:0706.2191}}].

\bibitem{Andronic:2009gj}
A.~Andronic et~al., {\it {Hadron Production in Ultra-relativistic Nuclear
  Collisions: Quarkyonic Matter and a Triple Point in the Phase Diagram of
  QCD}},  {\em Nucl. Phys. A} {\bf 837} (2010) 65--86,
  [\href{http://arxiv.org/abs/0911.4806}{{\tt arXiv:0911.4806}}].

\bibitem{Fukushima:2013rx}
K.~Fukushima and C.~Sasaki, {\it {The phase diagram of nuclear and quark matter
  at high baryon density}},  {\em Prog. Part. Nucl. Phys.} {\bf 72} (2013)
  99--154, [\href{http://arxiv.org/abs/1301.6377}{{\tt arXiv:1301.6377}}].

\bibitem{Philipsen:2019qqm}
O.~Philipsen and J.~Scheunert, {\it {QCD in the heavy dense regime for general
  $N_c$: On the existence of quarkyonic matter}},  {\em JHEP} {\bf 11} (2019)
  022, [\href{http://arxiv.org/abs/1908.03136}{{\tt arXiv:1908.03136}}].

\bibitem{Fukushima:2008wg}
K.~Fukushima, {\it {Phase diagrams in the three-flavor Nambu-Jona-Lasinio model
  with the Polyakov loop}},  {\em Phys. Rev. D} {\bf 77} (2008) 114028,
  [\href{http://arxiv.org/abs/0803.3318}{{\tt arXiv:0803.3318}}]. [Erratum:
  Phys.Rev.D 78, 039902 (2008)].

\bibitem{McLerran:2008ua}
L.~McLerran, K.~Redlich, and C.~Sasaki, {\it {Quarkyonic Matter and Chiral
  Symmetry Breaking}},  {\em Nucl. Phys. A} {\bf 824} (2009) 86--100,
  [\href{http://arxiv.org/abs/0812.3585}{{\tt arXiv:0812.3585}}].

\bibitem{Sakai:2011fa}
Y.~Sakai, T.~Sasaki, H.~Kouno, and M.~Yahiro, {\it {Equation of state in the
  PNJL model with the entanglement interaction}},  {\em J. Phys. G} {\bf 39}
  (2012) 035004, [\href{http://arxiv.org/abs/1104.2394}{{\tt
  arXiv:1104.2394}}].

\bibitem{Kojo:2009ha}
T.~Kojo, Y.~Hidaka, L.~McLerran, and R.~D. Pisarski, {\it {Quarkyonic Chiral
  Spirals}},  {\em Nucl. Phys. A} {\bf 843} (2010) 37--58,
  [\href{http://arxiv.org/abs/0912.3800}{{\tt arXiv:0912.3800}}].

\bibitem{Kojo:2011cn}
T.~Kojo, Y.~Hidaka, K.~Fukushima, L.~D. McLerran, and R.~D. Pisarski, {\it
  {Interweaving Chiral Spirals}},  {\em Nucl. Phys. A} {\bf 875} (2012)
  94--138, [\href{http://arxiv.org/abs/1107.2124}{{\tt arXiv:1107.2124}}].

\bibitem{McLerran:2018hbz}
L.~McLerran and S.~Reddy, {\it {Quarkyonic Matter and Neutron Stars}},  {\em
  Phys. Rev. Lett.} {\bf 122} (2019), no.~12 122701,
  [\href{http://arxiv.org/abs/1811.12503}{{\tt arXiv:1811.12503}}].

\bibitem{Jeong:2019lhv}
K.~S. Jeong, L.~McLerran, and S.~Sen, {\it {Dynamical Derivation of the
  Momentum Space Shell Structure for Quarkyonic Matter}},  {\em Phys. Rev. C}
  {\bf 101} (2020), no.~3 035201, [\href{http://arxiv.org/abs/1908.04799}{{\tt
  arXiv:1908.04799}}].

\bibitem{Sen:2020peq}
S.~Sen and N.~C. Warrington, {\it {Finite-Temperature Quarkyonic Matter with an
  Excluded Volume Model for Nuclear Interactions}},
  \href{http://arxiv.org/abs/2002.11133}{{\tt arXiv:2002.11133}}.

\bibitem{Duarte:2020xsp}
D.~C. Duarte, S.~Hernandez-Ortiz, and K.~S. Jeong, {\it {Excluded Volume Model
  for Quarkyonic Matter: 3-Flavor Baryon-Quark Mixture}},
  \href{http://arxiv.org/abs/2003.02362}{{\tt arXiv:2003.02362}}.

\bibitem{Zhao:2020dvu}
T.~Zhao and J.~M. Lattimer, {\it {Quarkyonic Matter Equation of State in
  Beta-Equilibrium}},  \href{http://arxiv.org/abs/2004.08293}{{\tt
  arXiv:2004.08293}}.

\bibitem{Witten:1998zw}
E.~Witten, {\it {Anti-de Sitter space, thermal phase transition, and
  confinement in gauge theories}},  {\em Adv.Theor.Math.Phys.} {\bf 2} (1998)
  505--532, [\href{http://arxiv.org/abs/hep-th/9803131}{{\tt hep-th/9803131}}].

\bibitem{Sakai:2004cn}
T.~Sakai and S.~Sugimoto, {\it {Low energy hadron physics in holographic QCD}},
   {\em Prog. Theor. Phys.} {\bf 113} (2005) 843--882,
  [\href{http://arxiv.org/abs/hep-th/0412141}{{\tt hep-th/0412141}}].

\bibitem{Sakai:2005yt}
T.~Sakai and S.~Sugimoto, {\it {More on a holographic dual of QCD}},  {\em
  Prog. Theor. Phys.} {\bf 114} (2005) 1083--1118,
  [\href{http://arxiv.org/abs/hep-th/0507073}{{\tt hep-th/0507073}}].

\bibitem{Bergman:2007wp}
O.~Bergman, G.~Lifschytz, and M.~Lippert, {\it {Holographic Nuclear Physics}},
  {\em JHEP} {\bf 11} (2007) 056, [\href{http://arxiv.org/abs/0708.0326}{{\tt
  arXiv:0708.0326}}].

\bibitem{Kovensky:2019bih}
N.~Kovensky and A.~Schmitt, {\it {Heavy Holographic QCD}},  {\em JHEP} {\bf 02}
  (2020) 096, [\href{http://arxiv.org/abs/1911.08433}{{\tt arXiv:1911.08433}}].

\bibitem{Aharony:2006da}
O.~Aharony, J.~Sonnenschein, and S.~Yankielowicz, {\it {A Holographic model of
  deconfinement and chiral symmetry restoration}},  {\em Annals Phys.} {\bf
  322} (2007) 1420--1443, [\href{http://arxiv.org/abs/hep-th/0604161}{{\tt
  hep-th/0604161}}].

\bibitem{Horigome:2006xu}
N.~Horigome and Y.~Tanii, {\it {Holographic chiral phase transition with
  chemical potential}},  {\em JHEP} {\bf 01} (2007) 072,
  [\href{http://arxiv.org/abs/hep-th/0608198}{{\tt hep-th/0608198}}].

\bibitem{Ghoroku:2012am}
K.~Ghoroku, K.~Kubo, M.~Tachibana, T.~Taminato, and F.~Toyoda, {\it
  {Holographic cold nuclear matter as dilute instanton gas}},  {\em Phys. Rev.
  D} {\bf 87} (2013), no.~6 066006, [\href{http://arxiv.org/abs/1211.2499}{{\tt
  arXiv:1211.2499}}].

\bibitem{Li:2015uea}
S.-w. Li, A.~Schmitt, and Q.~Wang, {\it {From holography towards real-world
  nuclear matter}},  {\em Phys. Rev.} {\bf D92} (2015), no.~2 026006,
  [\href{http://arxiv.org/abs/1505.04886}{{\tt arXiv:1505.04886}}].

\bibitem{Preis:2016fsp}
F.~Preis and A.~Schmitt, {\it {Layers of deformed instantons in holographic
  baryonic matter}},  {\em JHEP} {\bf 07} (2016) 001,
  [\href{http://arxiv.org/abs/1606.00675}{{\tt arXiv:1606.00675}}].

\bibitem{BitaghsirFadafan:2018uzs}
K.~Bitaghsir~Fadafan, F.~Kazemian, and A.~Schmitt, {\it {Towards a holographic
  quark-hadron continuity}},  {\em JHEP} {\bf 03} (2019) 183,
  [\href{http://arxiv.org/abs/1811.08698}{{\tt arXiv:1811.08698}}].

\bibitem{Kaplunovsky:2012gb}
V.~Kaplunovsky, D.~Melnikov, and J.~Sonnenschein, {\it {Baryonic Popcorn}},
  {\em JHEP} {\bf 11} (2012) 047, [\href{http://arxiv.org/abs/1201.1331}{{\tt
  arXiv:1201.1331}}].

\bibitem{Kaplunovsky:2013iza}
V.~Kaplunovsky and J.~Sonnenschein, {\it {Dimension Changing Phase Transitions
  in Instanton Crystals}},  {\em JHEP} {\bf 04} (2014) 022,
  [\href{http://arxiv.org/abs/1304.7540}{{\tt arXiv:1304.7540}}].

\bibitem{deBoer:2012ij}
J.~de~Boer, B.~D. Chowdhury, M.~P. Heller, and J.~Jankowski, {\it {Towards a
  holographic realization of the Quarkyonic phase}},  {\em Phys. Rev. D} {\bf
  87} (2013), no.~6 066009, [\href{http://arxiv.org/abs/1209.5915}{{\tt
  arXiv:1209.5915}}].

\bibitem{Chen:2019rez}
X.~Chen, D.~Li, D.~Hou, and M.~Huang, {\it {Quarkyonic phase from quenched
  dynamical holographic QCD model}},  {\em JHEP} {\bf 03} (2020) 073,
  [\href{http://arxiv.org/abs/1908.02000}{{\tt arXiv:1908.02000}}].

\bibitem{Ishii:2019gta}
T.~Ishii, M.~J{\"a}rvinen, and G.~Nijs, {\it {Cool baryon and quark matter in
  holographic QCD}},  {\em JHEP} {\bf 07} (2019) 003,
  [\href{http://arxiv.org/abs/1903.06169}{{\tt arXiv:1903.06169}}].

\bibitem{Schafer:1998ef}
T.~Sch{\"a}fer and F.~Wilczek, {\it {Continuity of quark and hadron matter}},
  {\em Phys. Rev. Lett.} {\bf 82} (1999) 3956--3959,
  [\href{http://arxiv.org/abs/hep-ph/9811473}{{\tt hep-ph/9811473}}].

\bibitem{Hatsuda:2006ps}
T.~Hatsuda, M.~Tachibana, N.~Yamamoto, and G.~Baym, {\it {New critical point
  induced by the axial anomaly in dense QCD}},  {\em Phys. Rev. Lett.} {\bf 97}
  (2006) 122001, [\href{http://arxiv.org/abs/hep-ph/0605018}{{\tt
  hep-ph/0605018}}].

\bibitem{Schmitt:2010pf}
A.~Schmitt, S.~Stetina, and M.~Tachibana, {\it {Ginzburg-Landau phase diagram
  for dense matter with axial anomaly, strange quark mass, and meson
  condensation}},  {\em Phys. Rev.} {\bf D83} (2011) 045008,
  [\href{http://arxiv.org/abs/1010.4243}{{\tt arXiv:1010.4243}}].

\bibitem{Baym:2019iky}
G.~Baym, S.~Furusawa, T.~Hatsuda, T.~Kojo, and H.~Togashi, {\it {New Neutron
  Star Equation of State with Quark-Hadron Crossover}},  {\em Astrophys. J.}
  {\bf 885} (2019) 42, [\href{http://arxiv.org/abs/1903.08963}{{\tt
  arXiv:1903.08963}}].

\bibitem{Preis:2010cq}
F.~Preis, A.~Rebhan, and A.~Schmitt, {\it {Inverse magnetic catalysis in dense
  holographic matter}},  {\em JHEP} {\bf 1103} (2011) 033,
  [\href{http://arxiv.org/abs/1012.4785}{{\tt arXiv:1012.4785}}].

\bibitem{Preis:2012fh}
F.~Preis, A.~Rebhan, and A.~Schmitt, {\it {Inverse magnetic catalysis in field
  theory and gauge-gravity duality}},  {\em Lect.Notes Phys.} {\bf 871} (2013)
  51--86, [\href{http://arxiv.org/abs/1208.0536}{{\tt arXiv:1208.0536}}].

\bibitem{Antonyan:2006vw}
E.~Antonyan, J.~A. Harvey, S.~Jensen, and D.~Kutasov, {\it {NJL and QCD from
  string theory}},  \href{http://arxiv.org/abs/hep-th/0604017}{{\tt
  hep-th/0604017}}.

\bibitem{Davis:2007ka}
J.~L. Davis, M.~Gutperle, P.~Kraus, and I.~Sachs, {\it {Stringy NJL and
  Gross-Neveu models at finite density and temperature}},  {\em JHEP} {\bf
  0710} (2007) 049, [\href{http://arxiv.org/abs/0708.0589}{{\tt
  arXiv:0708.0589}}].

\bibitem{Rebhan:2014rxa}
A.~Rebhan, {\it {The Witten-Sakai-Sugimoto model: A brief review and some
  recent results}},  {\em EPJ Web Conf.} {\bf 95} (2015) 02005,
  [\href{http://arxiv.org/abs/1410.8858}{{\tt arXiv:1410.8858}}].

\bibitem{Aharony:2008an}
O.~Aharony and D.~Kutasov, {\it {Holographic Duals of Long Open Strings}},
  {\em Phys. Rev.} {\bf D78} (2008) 026005,
  [\href{http://arxiv.org/abs/0803.3547}{{\tt arXiv:0803.3547}}].

\bibitem{Argyres:2008sw}
P.~C. Argyres, M.~Edalati, R.~G. Leigh, and J.~F. Vazquez-Poritz, {\it {Open
  Wilson Lines and Chiral Condensates in Thermal Holographic QCD}},  {\em Phys.
  Rev.} {\bf D79} (2009) 045022, [\href{http://arxiv.org/abs/0811.4617}{{\tt
  arXiv:0811.4617}}].

\bibitem{Hata:2007mb}
H.~Hata, T.~Sakai, S.~Sugimoto, and S.~Yamato, {\it {Baryons from instantons in
  holographic QCD}},  {\em Prog. Theor. Phys.} {\bf 117} (2007) 1157,
  [\href{http://arxiv.org/abs/hep-th/0701280}{{\tt hep-th/0701280}}].

\bibitem{Seki:2008mu}
S.~Seki and J.~Sonnenschein, {\it {Comments on Baryons in Holographic QCD}},
  {\em JHEP} {\bf 01} (2009) 053, [\href{http://arxiv.org/abs/0810.1633}{{\tt
  arXiv:0810.1633}}].

\bibitem{Preis:2011sp}
F.~Preis, A.~Rebhan, and A.~Schmitt, {\it {Holographic baryonic matter in a
  background magnetic field}},  {\em J. Phys. G} {\bf 39} (2012) 054006,
  [\href{http://arxiv.org/abs/1109.6904}{{\tt arXiv:1109.6904}}].

\bibitem{Elliot-Ripley:2016uwb}
M.~Elliot-Ripley, P.~Sutcliffe, and M.~Zamaklar, {\it {Phases of kinky
  holographic nuclear matter}},  {\em JHEP} {\bf 10} (2016) 088,
  [\href{http://arxiv.org/abs/1607.04832}{{\tt arXiv:1607.04832}}].

\bibitem{Hashimoto:2008sr}
K.~Hashimoto, T.~Hirayama, F.-L. Lin, and H.-U. Yee, {\it {Quark Mass
  Deformation of Holographic Massless QCD}},  {\em JHEP} {\bf 07} (2008) 089,
  [\href{http://arxiv.org/abs/0803.4192}{{\tt arXiv:0803.4192}}].

\bibitem{McNees:2008km}
R.~McNees, R.~C. Myers, and A.~Sinha, {\it {On quark masses in holographic
  QCD}},  {\em JHEP} {\bf 11} (2008) 056,
  [\href{http://arxiv.org/abs/0807.5127}{{\tt arXiv:0807.5127}}].

\bibitem{Callebaut:2011ab}
N.~Callebaut, D.~Dudal, and H.~Verschelde, {\it {Holographic rho mesons in an
  external magnetic field}},  {\em JHEP} {\bf 03} (2013) 033,
  [\href{http://arxiv.org/abs/1105.2217}{{\tt arXiv:1105.2217}}].

\bibitem{1975PhLB...59...85B}
A.~A. {Belavin}, A.~M. {Polyakov}, A.~S. {Schwartz}, and Y.~S. {Tyupkin}, {\it
  {Pseudoparticle solutions of the Yang-Mills equations}},  {\em Physics
  Letters B} {\bf 59} (Oct., 1975) 85--87.

\bibitem{Hashimoto:2009hj}
K.~Hashimoto, T.~Hirayama, and D.~K. Hong, {\it {Quark Mass Dependence of
  Hadron Spectrum in Holographic QCD}},  {\em Phys. Rev.} {\bf D81} (2010)
  045016, [\href{http://arxiv.org/abs/0906.0402}{{\tt arXiv:0906.0402}}].

\bibitem{Karch:2008fa}
A.~Karch, D.~Son, and A.~Starinets, {\it {Zero Sound from Holography}},
  \href{http://arxiv.org/abs/0806.3796}{{\tt arXiv:0806.3796}}.

\bibitem{Kulaxizi:2008jx}
M.~Kulaxizi and A.~Parnachev, {\it {Holographic Responses of Fermion Matter}},
  {\em Nucl. Phys. B} {\bf 815} (2009) 125--141,
  [\href{http://arxiv.org/abs/0811.2262}{{\tt arXiv:0811.2262}}].

\bibitem{DiNunno:2014bxa}
B.~S. DiNunno, M.~Ihl, N.~Jokela, and J.~F. Pedraza, {\it {Holographic zero
  sound at finite temperature in the Sakai-Sugimoto model}},  {\em JHEP} {\bf
  04} (2014) 149, [\href{http://arxiv.org/abs/1403.1827}{{\tt
  arXiv:1403.1827}}].

\bibitem{Glendenning:1992vb}
N.~K. Glendenning, {\it {First order phase transitions with more than one
  conserved charge: Consequences for neutron stars}},  {\em Phys. Rev.} {\bf
  D46} (1992) 1274--1287.

\bibitem{Heiselberg:1992dx}
H.~Heiselberg, C.~J. Pethick, and E.~F. Staubo, {\it {Quark matter droplets in
  neutron stars}},  {\em Phys. Rev. Lett.} {\bf 70} (1993) 1355--1359.

\bibitem{Schmitt:2020tac}
A.~Schmitt, {\it {Chiral pasta: Mixed phases at the chiral phase transition}},
  {\em Phys. Rev. D} {\bf 101} (2020), no.~7 074007,
  [\href{http://arxiv.org/abs/2002.01451}{{\tt arXiv:2002.01451}}].

\bibitem{Mateos:2007vn}
D.~Mateos, R.~C. Myers, and R.~M. Thomson, {\it {Thermodynamics of the brane}},
   {\em JHEP} {\bf 05} (2007) 067,
  [\href{http://arxiv.org/abs/hep-th/0701132}{{\tt hep-th/0701132}}].

\bibitem{Rozali:2007rx}
M.~Rozali, H.-H. Shieh, M.~Van~Raamsdonk, and J.~Wu, {\it {Cold Nuclear Matter
  In Holographic QCD}},  {\em JHEP} {\bf 01} (2008) 053,
  [\href{http://arxiv.org/abs/0708.1322}{{\tt arXiv:0708.1322}}].

\bibitem{Most:2019onn}
E.~R. Most, L.~Jens~Papenfort, V.~Dexheimer, M.~Hanauske, H.~Stoecker, and
  L.~Rezzolla, {\it {On the deconfinement phase transition in neutron-star
  mergers}},  {\em Eur. Phys. J. A} {\bf 56} (2020), no.~2 59,
  [\href{http://arxiv.org/abs/1910.13893}{{\tt arXiv:1910.13893}}].

\bibitem{Dexheimer:2009hi}
V.~Dexheimer and S.~Schramm, {\it {A Novel Approach to Model Hybrid Stars}},
  {\em Phys. Rev. C} {\bf 81} (2010) 045201,
  [\href{http://arxiv.org/abs/0901.1748}{{\tt arXiv:0901.1748}}].

\bibitem{Aryal:2020ocm}
K.~Aryal, C.~Constantinou, R.~Farias, and V.~Dexheimer, {\it {QCD Phase
  Diagrams with Charge and Isospin Axes under Heavy-Ion Collision and Stellar
  Conditions}},  \href{http://arxiv.org/abs/2004.03039}{{\tt
  arXiv:2004.03039}}.

\bibitem{Marczenko:2020jma}
M.~Marczenko, D.~Blaschke, K.~Redlich, and C.~Sasaki, {\it {Towards a unified
  equation of state for multi-messenger astronomy}},
  \href{http://arxiv.org/abs/2004.09566}{{\tt arXiv:2004.09566}}.

\bibitem{Fukushima:2015bda}
K.~Fukushima and T.~Kojo, {\it {The Quarkyonic Star}},  {\em Astrophys. J.}
  {\bf 817} (2016), no.~2 180, [\href{http://arxiv.org/abs/1509.00356}{{\tt
  arXiv:1509.00356}}].

\bibitem{BitaghsirFadafan:2018iqr}
K.~Bitaghsir~Fadafan, J.~Cruz~Rojas, and N.~Evans, {\it {Holographic
  description of color superconductivity}},  {\em Phys. Rev. D} {\bf 98}
  (2018), no.~6 066010, [\href{http://arxiv.org/abs/1803.03107}{{\tt
  arXiv:1803.03107}}].

\bibitem{Rebhan:2008ur}
A.~Rebhan, A.~Schmitt, and S.~A. Stricker, {\it {Meson supercurrents and the
  Meissner effect in the Sakai-Sugimoto model}},  {\em JHEP} {\bf 05} (2009)
  084, [\href{http://arxiv.org/abs/0811.3533}{{\tt arXiv:0811.3533}}].

\bibitem{BallonBayona:2012wx}
A.~Ballon-Bayona, K.~Peeters, and M.~Zamaklar, {\it {A chiral magnetic spiral
  in the holographic Sakai-Sugimoto model}},  {\em JHEP} {\bf 11} (2012) 164,
  [\href{http://arxiv.org/abs/1209.1953}{{\tt arXiv:1209.1953}}].

\bibitem{Bigazzi:2014qsa}
F.~Bigazzi and A.~L. Cotrone, {\it {Holographic QCD with Dynamical Flavors}},
  {\em JHEP} {\bf 01} (2015) 104, [\href{http://arxiv.org/abs/1410.2443}{{\tt
  arXiv:1410.2443}}].

\bibitem{Li:2016gtz}
S.-w. Li and T.~Jia, {\it {Dynamically flavored description of holographic QCD
  in the presence of a magnetic field}},  {\em Phys. Rev.} {\bf D96} (2017),
  no.~6 066032, [\href{http://arxiv.org/abs/1604.07197}{{\tt
  arXiv:1604.07197}}].

\bibitem{Wagenbrunn:2007ie}
R.~Wagenbrunn and L.~Glozman, {\it {Chiral symmetry patterns of excited mesons
  with the Coulomb-like linear confinement}},  {\em Phys. Rev. D} {\bf 75}
  (2007) 036007, [\href{http://arxiv.org/abs/hep-ph/0701039}{{\tt
  hep-ph/0701039}}].

\bibitem{Glozman:2007tv}
L.~Glozman and R.~Wagenbrunn, {\it {Chirally symmetric but confining dense and
  cold matter}},  {\em Phys. Rev. D} {\bf 77} (2008) 054027,
  [\href{http://arxiv.org/abs/0709.3080}{{\tt arXiv:0709.3080}}].

\bibitem{Glozman:2012fj}
L.~Glozman, C.~Lang, and M.~Schrock, {\it {Symmetries of hadrons after
  unbreaking the chiral symmetry}},  {\em Phys. Rev. D} {\bf 86} (2012) 014507,
  [\href{http://arxiv.org/abs/1205.4887}{{\tt arXiv:1205.4887}}].

\bibitem{Hirayama:2019vod}
K.~Zhang, T.~Hirayama, L.-W. Luo, and F.-L. Lin, {\it {Compact Star of
  Holographic Nuclear Matter and GW170817}},  {\em Phys. Lett. B} {\bf 801}
  (2020) 135176, [\href{http://arxiv.org/abs/1902.08477}{{\tt
  arXiv:1902.08477}}].

\bibitem{Hoyos:2016zke}
C.~Hoyos, D.~Rodr{\'i}guez~Fern{\'a}ndez, N.~Jokela, and A.~Vuorinen, {\it
  {Holographic quark matter and neutron stars}},  {\em Phys. Rev. Lett.} {\bf
  117} (2016), no.~3 032501, [\href{http://arxiv.org/abs/1603.02943}{{\tt
  arXiv:1603.02943}}].

\bibitem{Chesler:2019osn}
P.~M. Chesler, N.~Jokela, A.~Loeb, and A.~Vuorinen, {\it {Finite-temperature
  Equations of State for Neutron Star Mergers}},  {\em Phys. Rev. D} {\bf 100}
  (2019), no.~6 066027, [\href{http://arxiv.org/abs/1906.08440}{{\tt
  arXiv:1906.08440}}].

\bibitem{Ecker:2019xrw}
C.~Ecker, M.~J{\"a}rvinen, G.~Nijs, and W.~van~der Schee, {\it {Gravitational
  waves from holographic neutron star mergers}},  {\em Phys. Rev. D} {\bf 101}
  (2020), no.~10 103006, [\href{http://arxiv.org/abs/1908.03213}{{\tt
  arXiv:1908.03213}}].

\bibitem{Fadafa:2019euu}
K.~Bitaghsir~Fadafan, J.~Cruz~Rojas, and N.~Evans, {\it {Deconfined, Massive
  Quark Phase at High Density and Compact Stars: A Holographic Study}},  {\em
  Phys. Rev. D} {\bf 101} (2020), no.~12 126005,
  [\href{http://arxiv.org/abs/1911.12705}{{\tt arXiv:1911.12705}}].

\bibitem{Jokela:2020piw}
N.~Jokela, M.~J{\"a}rvinen, G.~Nijs, and J.~Remes, {\it {Unified weak/strong
  coupling framework for nuclear matter and neutron stars}},
  \href{http://arxiv.org/abs/2006.01141}{{\tt arXiv:2006.01141}}.

\end{thebibliography}\endgroup

\end{document}